\documentclass[12pt,english]{article}

\usepackage{mathptmx}
\usepackage{geometry}
\usepackage[fleqn]{amsmath}
\usepackage{pifont}
\newcommand{\cmark}{\ding{51}}%
\newcommand{\xmark}{\ding{55}}%
\usepackage{xcolor}
\definecolor{mygray}{gray}{0.6}
\usepackage{algorithm}
\usepackage{algorithmic}

\newcommand{\setR}{\mathbb{R}}
\DeclareMathOperator*{\argmin}{arg\,min}

\usepackage{array}
\usepackage{bbding}
\usepackage{multirow}
\usepackage{amssymb}
\usepackage{amsthm}
\usepackage{graphicx}
\usepackage{natbib}
\usepackage{lscape}
\usepackage[hyphens]{url}
\usepackage[hidelinks]{hyperref}
\usepackage{comment}
\usepackage{bm}
\usepackage[title]{appendix}
\usepackage{longtable}
\usepackage{mathtools}
\usepackage{chngcntr}
\usepackage{authblk}
\usepackage{babel}
\usepackage{dsfont}
\usepackage{fancyhdr}
\usepackage{cleveref}
\usepackage{abstract}

\usepackage{stackengine}
\usepackage{textcomp}
\usepackage{stfloats}
\usepackage{verbatim}
\usepackage{xcolor}
\usepackage{microtype}
\usepackage{graphicx}
\usepackage{subfigure}
\usepackage{booktabs} 
\usepackage{bbm}

\usepackage{amsfonts}
\usepackage{xurl}
\usepackage{stackengine}
\usepackage{tikz}
\usetikzlibrary{decorations.pathreplacing}
\usetikzlibrary{positioning,arrows.meta,quotes}
\usetikzlibrary{shapes,snakes}
\usetikzlibrary{bayesnet}
\tikzset{>=latex}
\tikzstyle{plate caption} = [caption, node distance=0, inner sep=0pt, below left=5pt and 0pt of #1.south]
\usepackage[normalem]{ulem}
\usepackage{multirow}\usetikzlibrary{positioning}

\usepackage{makecell}
\makeatletter
\allowdisplaybreaks
\date{}

\fancypagestyle{plain}{%
	\fancyhf{} 
	
	\lhead{\footnotesize \textcolor{gray}{arXiv preprint}}
	\chead{\footnotesize \textcolor{gray}{\textit{\shortauthors -- \runningtitle}}}
	\rhead{\footnotesize \textcolor{gray}{\thepage}}
}
\makeatletter
\def\blfootnote{\xdef\@thefnmark{}\@footnotetext}
\makeatother

\makeatletter
\def\titlepageext{
	\begin{center}	
	{\parindent0pt
		\rule{0.9\textwidth}{1pt}
		\begin{minipage}[t]{0.25\textwidth}
			\small {\it Keywords:}\\
			\keyword
		\end{minipage}%
		\hspace{3mm}
		\begin{minipage}[t]{0.6\textwidth}
			\small \abstract
		\end{minipage}%
		
		\rule{0.9\textwidth}{2pt}
	}
	\end{center}

	\blfootnote{* Corresponding author. E-mail address: \href{mailto:\corresemail}{\corresemail}.}
}
\makeatother

\usepackage[mathlines, pagewise]{lineno}
\usepackage{etoolbox} 

\newcommand*\linenomathpatchAMS[1]{%
	\expandafter\pretocmd\csname #1\endcsname {\linenomathAMS}{}{}%
	\expandafter\pretocmd\csname #1*\endcsname{\linenomathAMS}{}{}%
	\expandafter\apptocmd\csname end#1\endcsname {\endlinenomath}{}{}%
	\expandafter\apptocmd\csname end#1*\endcsname{\endlinenomath}{}{}%
}

\expandafter\ifx\linenomath\linenomathWithnumbers
\let\linenomathAMS\linenomathWithnumbers
\patchcmd\linenomathAMS{\advance\postdisplaypenalty\linenopenalty}{}{}{}
\else
\let\linenomathAMS\linenomathNonumbers
\fi

\linenomathpatchAMS{gather}
\linenomathpatchAMS{multline}
\linenomathpatchAMS{align}
\linenomathpatchAMS{alignat}
\linenomathpatchAMS{flalign}

\nolinenumbers

\title{When Context Is Not Enough: Modeling Unexplained\\ Variability in Car-Following Behavior}

\def\shortauthors{Zhang et al.}
\def\runningtitle{Modeling Unexplained Variability in Car-Following Behavior.}

\author[a]{Chengyuan Zhang}
\author[b]{Zhengbing He}
\author[b]{Cathy Wu}
\author[a,$\ast$]{Lijun Sun}
\affil[a]{Department of Civil Engineering, McGill University, 817 Sherbrooke Street West, Montreal, QC  H3A 0C3, Canada}
\affil[b]{Laboratory for Information \& Decision Systems (LIDS), Massachusetts Institute of Technology, Cambridge, MA 02139, USA}

\def\corresemail{lijun.sun@mcgill.ca}

\def\abstract{Modeling car-following behavior is fundamental to microscopic traffic simulation, yet traditional deterministic models often fail to capture the full extent of variability and unpredictability in human driving. While many modern approaches incorporate context-aware inputs (e.g., spacing, speed, relative speed), they frequently overlook structured stochasticity that arises from latent driver intentions, perception errors, and memory effects---factors that are not directly observable from context alone. To fill the gap, this study introduces an interpretable stochastic modeling framework that captures not only context-dependent dynamics but also residual variability beyond what context can explain. Leveraging deep neural networks integrated with nonstationary Gaussian processes (GPs), our model employs a scenario-adaptive Gibbs kernel to learn dynamic temporal correlations in acceleration decisions, where the strength and duration of correlations between acceleration decisions evolve with the driving context. This formulation enables a principled, data-driven quantification of uncertainty in acceleration, speed, and spacing, grounded in both observable context and latent behavioral variability. Comprehensive experiments on the naturalistic vehicle trajectory dataset collected from the German highway, i.e., the HighD dataset, demonstrate that the proposed stochastic simulation method within this framework surpasses conventional methods in both predictive performance and interpretable uncertainty quantification. The integration of interpretability and accuracy makes this framework a promising tool for traffic analysis and safety-critical applications.
}

\def\keyword{car-following behavior\\
stochastic simulation\\
temporal correlation\\
Gaussian processes\\
interpretability}

\begin{document}
\maketitle
\titlepageext

\newpage

\section{Introduction}

Modeling and simulating car-following behaviors is a cornerstone of microscopic transportation studies and practice, directly influencing the fidelity of traffic simulation and the effectiveness of transportation control and optimization. Traditional car-following models, such as the Intelligent Driver Model (IDM) \citep{treiber2000congested} and Gipps' model \citep{gipps1981behavioural}, typically use deterministic rules to represent interactions between leading and following vehicles. While these models capture key behavioral mechanisms, they often treat human responses as deterministic or introduce only simplistic, temporally uncorrelated disturbances. {Such assumptions are limiting because driver actions are shaped not only by observable kinematic states but also by unobserved and time-varying factors, including memory and adaptation, perception uncertainty, and latent situational cues (e.g., brake-light signals, weather and visibility, road geometry, and driver internal states) that are rarely available in standard trajectory datasets \citep{wang2019long}. Consequently, even when conditioning on common car-following stimuli (e.g., spacing and relative speed), the remaining discrepancy typically exhibits persistent stochastic components and regime-dependent variability rather than behaving as i.i.d.\ noise. This view is supported by trajectory-data and experimental evidence and has motivated stochastic car-following formulations that model acceleration or speed fluctuations as random processes, demonstrating that structured stochasticity can be essential for reproducing observed oscillations and variability in traffic dynamics \citep{laval2014parsimonious,ngoduy2019langevin,xu2020statistical}.}

Recent studies have introduced various stochastic car-following models that explicitly incorporate randomness to capture the variability in human driving behavior.
For instance, the stochastic extensions to deterministic models \citep{treiber2006delays, zhang2024calibrating} utilize processes like Ornstein--Uhlenbeck (OU) or autoregressive (AR) models to model fluctuations in acceleration, speed, and spacing. These approaches introduce stochastic noise terms into otherwise deterministic dynamics, enabling the representation of uncertainty in human drivers' responses. However, because these methods typically rely on stationary or linear assumptions, their ability to capture temporal dynamics may be limited, especially in complex driving scenarios where behavioral patterns are nonlinear, context-dependent, or rapidly evolving. 
For example, drivers may exhibit delayed or anticipatory responses, variable reaction intensities depending on traffic states, or adaptive behaviors that change abruptly during lane changes or sudden braking events. These dynamics give rise to residual stochasticity that reflects not only context but also unobserved intent, perception errors, and memory effects---factors that traditional stochastic models generally overlook \citep{zhang2024bayesian, treiber2006delays, bifulco2013driving, wang2020stability}.
Motivated by these challenges, recent work has turned to deep learning methods, particularly recurrent neural networks (RNNs) and their variants \citep{morton2016analysis, zhou2017recurrent, wang2017capturing}, which have shown strong empirical performance in learning car-following patterns directly from trajectory data. These models adapt flexibly to diverse traffic contexts and behavioral styles, yet often lack explicit probabilistic structures for quantifying uncertainty or revealing the temporal dependencies in driver decision-making. As a result, it becomes difficult to interpret or diagnose \textit{why} certain driving actions occur, or \textit{how} prediction reliability evolves over time. This motivates the need for hybrid approaches that combine the representational power of deep models with structured components that offer interpretability and probabilistic insight.

In summary, real-world driving behavior exhibits complex and critical stochastic characteristics: drivers adapt their actions based on imperfect perception, varying attention, and latent intent, all of which evolve over time and across traffic contexts. These \textit{context-dependent stochastic features} are neither purely random nor stationary. Instead, they are structured, temporally correlated patterns that simple noise models cannot fully capture. Much of the existing literature, even when incorporating additive noise, does not account for this structure, resulting in simulations that are often unrealistic or poorly calibrated.

This work proposes an interpretable stochastic car-following framework that integrates deep neural sequence modeling with nonstationary GPs. {Using DeepAR \citep{salinas2020deepar} as one instantiation of the mean dynamics,} our model predicts both the mean acceleration and a context-dependent covariance structure via a scenario-adaptive Gibbs kernel. At each time step, the recurrent network ingests spacing, speed, and relative speed to output a predicted acceleration, a lengthscale, and a variance. These kernel parameters modulate a Gaussian-process residual that captures temporally correlated deviations from the mean dynamics, allowing the model to adapt its memory and uncertainty to evolving traffic conditions in an end-to-end trainable manner.

{The primary contribution of this work is the development of a stochastic modeling framework for car-following behavior that explicitly captures context-dependent temporal correlations in human driving. Conceptually, we distinguish \emph{aleatoric} uncertainty arising from irreducible variability in driving actions and sensing/measurement noise from \emph{epistemic} uncertainty due to limited coverage of driving contexts or model misspecification. Our framework focuses on representing structured aleatoric variability through a temporally correlated residual process whose correlation and variance adapt to driving context, while providing a clear lens for interpreting potential epistemic effects under domain shift or sparse contexts.} This core modeling innovation delivers three key benefits. First, it provides a principled way to represent human driving variability as a temporally correlated stochastic process, overcoming the stationarity assumptions of classical models. Second, the scenario-adaptive Gibbs kernel enables the model to learn dynamic correlation scales and heteroskedastic noise levels that reflect different driving regimes in a data-driven and interpretable manner. Finally, extensive experiments on naturalistic HighD trajectories show that our approach yields lower simulation errors and sharper, better-calibrated uncertainty estimates than both deterministic baselines and stationary GP variants. Together, these advantages make the framework a powerful and interpretable tool for high-fidelity traffic simulation and safety-critical behavior analysis.

The remainder of this paper is organized as follows. 
\Cref{sec:related_work} reviews relevant literature on car-following models and stochastic simulations. 
\Cref{sec:methodology} presents the proposed stochastic modeling framework. \Cref{sec:exp_and_results} describes the experimental setup, stochastic simulation method, and evaluation results. \Cref{sec:conclusion} concludes the study and outlines directions for future work.

\section{Related Work} \label{sec:related_work}
Understanding and modeling car-following behavior has been a longstanding focus in transportation research. Over the decades, a wide range of models have been proposed, from rule-based formulations grounded in traffic flow theory to flexible machine learning approaches trained on naturalistic driving data \citep{zheng2023multi, he2022physics}. Comprehensive surveys such as \citet{chen2023follownet} and \citet{zhang2024car} provide detailed overviews of this evolving landscape. Building on this foundation, this section highlights two major lines of research most relevant to our study: (i) traditional analytical models, typically deterministic in nature, augmented with stochastic components to capture behavioral variability, and (ii) data-driven methods, particularly those leveraging deep learning and sequential modeling to learn complex driving patterns directly from observed trajectories.

\subsection{Traditional Analytical Models with Stochastic Extensions}

Traditional car-following models such as the General Motors (GM) stimulus–response model, the Optimal Velocity Model (OVM), and the IDM have played a foundational role in the development of microscopic traffic theories. The GM model links a follower vehicle’s acceleration to the relative speed and headway with respect to the leader, providing an intuitive yet simplistic response mechanism \citep{gazis1961nonlinear}. The OVM introduces a desired speed function that depends on headway, capturing emergent traffic phenomena such as spontaneous congestion waves and phase transitions between free-flow and stop-and-go conditions \citep{bando1995dynamical}. The IDM advances this framework by incorporating a driver's desired speed, time gap, and acceleration limits into a smooth and collision-avoiding formulation \citep{treiber2000congested}. { Beyond traditional stimulus-response frameworks, more sophisticated analytical models have been developed to account for human physiological and psychological constraints. For instance, psychophysical models, such as the Wiedemann model \citep{wiedemann1974simulation}, utilize action-point thresholds to describe discontinuities in acceleration, reflecting the limits of human perception in detecting small changes in relative speed or distance.
While driver heterogeneity can be incorporated in simulation practice (e.g., via driver classes or parameter distributions), many analytical car-following models are still often deployed with fixed or slowly varying parameterizations within a given run and with simplified disturbance assumptions (frequently i.i.d.\ and context-invariant). Even when multiple parameter sets are used across traffic states, the remaining fluctuations are rarely modeled as a \emph{context-dependent, temporally correlated} residual process. This motivates our framework, which complements existing mean-model formulations by learning regime-dependent variance and temporal correlation in the residual and can be integrated on top of heterogeneous parameterizations.}

In response, Bayesian methods have been widely adopted to enrich classical models with stochastic structure. By treating model parameters as random variables, Bayesian calibration enables the quantification of uncertainty and facilitates more robust fitting to observed trajectories. For instance, recent studies employ Bayesian dynamic regression and Markov Chain Monte Carlo (MCMC) techniques to infer posterior distributions over IDM parameters, thereby capturing inter-driver variability and enhancing predictive realism \citep{zhang2024calibrating}. Complementary approaches such as Approximate Bayesian Computation (ABC) have also shown promise, especially in scenarios where likelihood functions are intractable or unknown \citep{jiang2025stochastic}.

Beyond parameter uncertainty, a parallel line of research has leveraged latent-variable models to capture the multi-modality and nonstationarity of driving behavior. Hidden Markov Models (HMMs) and Gaussian Mixture Models (GMMs) are frequently used to represent discrete behavioral modes, such as accelerating, cruising, and braking, and to infer regime-switching dynamics from trajectory data \citep{zhang2025markov, vaitkus2014driving}. These models are particularly effective at reproducing realistic stop-and-go patterns and behavioral transitions. Related work on mixture models and latent class frameworks further extends this idea by probabilistically clustering driver types or behavioral styles \citep{chen2023bayesian, zhang2024learning}, offering a principled way to represent population-level heterogeneity.

More recently, GP models have emerged as a non-parametric Bayesian alternative for learning car-following dynamics directly from data. By modeling acceleration as a stochastic function of spacing, speed, and relative speed, GPs provide flexible representations with calibrated confidence intervals. Their ability to capture smooth, data-driven response surfaces makes them well-suited for modeling uncertainty in driver responses, although their use in traffic modeling has only recently gained traction \citep{soldevila2021car, wang2022gaussian}.

Taken together, these probabilistic approaches represent a significant departure from purely deterministic models, offering a richer and more realistic account of driving behavior by embedding uncertainty, heterogeneity, and latent structure into the modeling framework.

\subsection{Data-Driven and Neural Sequence Modeling Approaches}

In recent years, the development of large-scale trajectory datasets and advances in deep learning have fueled growing interest in using neural networks to model driving behavior directly from data \citep{Li2020aa}. Unlike traditional models that rely on manually crafted rules, data-driven approaches can learn complex, non-linear response patterns by optimizing over observed driving trajectories \citep{He2015a, huang2025survey}. Among these, recurrent neural networks (RNNs), particularly Long Short-Term Memory (LSTM) and Gated Recurrent Unit (GRU) architectures, have proven especially effective in capturing temporal dependencies inherent in car-following dynamics.

For instance, \citet{zhou2017recurrent} introduced a basic RNN-based car-following model that successfully reproduced traffic oscillation patterns within vehicle platoons. Building on this, LSTM-based models have shown enhanced capabilities in learning long-range temporal correlations by mitigating issues of vanishing gradients, making them suitable for representing delayed driver reactions and cumulative behavioral patterns \citep{ma2020sequence}. Empirical studies have consistently demonstrated that LSTM- and GRU-based models outperform classical car-following models in reproducing acceleration and braking behaviors, particularly in complex or congested traffic scenarios \citep{wang2017capturing, lin2020platoon}.
A further extension of these ideas is found in probabilistic sequence models, such as DeepAR \citep{salinas2020deepar}, which formulate driving behavior forecasting as an autoregressive sequence prediction problem. Rather than providing point estimates alone, DeepAR learns a full predictive distribution over future trajectories, thus offering both forecasts and associated uncertainty intervals. When trained on car-following sequences, these models can generate calibrated probabilistic forecasts that are crucial for risk-aware decision-making in autonomous driving systems. The ability to model uncertainty makes these approaches particularly well-suited for the inherently stochastic and context-sensitive nature of human driving.
Beyond recurrent models, several studies have benchmarked data-driven architectures, including feedforward neural networks, LSTMs, and reinforcement learning agents, against traditional physics-based models \citep{huang2021driving}. Results suggest that data-driven models can more accurately reproduce observed vehicle trajectories while better capturing nuanced behavioral patterns that are often missed by rule-based formulations \citep{zhu2018human}. These findings highlight the promise of modern neural architectures in learning rich representations of driving dynamics directly from data, thereby reducing reliance on hand-engineered assumptions.

More recently, hybrid modeling frameworks have emerged as a promising direction, aiming to bridge the gap between physical interpretability and data-driven flexibility. Rather than discarding classical models, these approaches integrate them into neural architectures as structured priors or corrective baselines. For example, \citet{10440530} proposed a neural-augmented IDM framework, where a neural network learns to compensate for discrepancies between the IDM and empirical behavior. Similarly, \citet{soldevila2021car} combined a parametric car-following model with a non-parametric GP module, allowing the model to learn data-driven corrections while preserving the theoretical structure of the base model. An interesting instance of hybridization is the use of time-varying parameters governed by latent embeddings, as demonstrated in \citet{zhang2022generative}, where the IDM is conditioned on a learned driver-style vector inferred through a neural process. This design allows the model to generalize across a range of behavioral regimes while maintaining interpretability. \textit{However, we note a common misconception in interpreting such frameworks: the parameters are often described as ``time-varying,'' when in fact, they are better understood as \textbf{context-dependent}. That is, the variation in parameters arises from the evolving driving state, including spacing, speed, and relative speed, rather than from an explicit dependence on time itself. Misinterpreting context-driven adaptation as purely temporal variation risks obscuring the structured and reactive nature of human driving behavior, and may lead to incorrect assumptions about model design and interpretation.} By embedding stochastic processes such as GPs or neural processes within deep architectures, these hybrid models achieve a balance between generalization and data efficiency. Grounding the learning process in well-established physical principles not only enhances interpretability but also helps ensure safety and constraint adherence, which is the critical aspects in traffic simulation and autonomous driving applications.

In summary, these developments reflect a growing consensus that hybrid, probabilistic, and deep learning approaches offer complementary strengths. However, existing methods still struggle to capture the nonstationary and context-dependent nature of temporal correlations in driving behavior, especially under dynamically changing traffic conditions. Many models either rely on stationary assumptions or do not explicitly model how uncertainty evolves over time. This work is motivated by the need to fill that gap by introducing a unified framework that integrates deep learning with nonstationary probabilistic modeling, enabling interpretable and high-fidelity simulation of human-like car-following behavior. The nonstationary formulations can capture evolving dynamics that vary with traffic context, scenario complexity, or driver state. Furthermore, models that expose internal structures, such as context-dependent lengthscales or uncertainty variances, offer interpretable insights into the decision-making regimes underlying human driving. This level of interpretability is especially valuable for behavior analysis, policy diagnostics, and uncertainty-aware traffic simulations in safety-critical contexts.

\section{Modeling Stochastic Car-Following Behavior with Nonstationary Temporal Correlations}\label{sec:methodology}
This section introduces a probabilistic framework for modeling and simulating car-following behaviors. Our goal is to capture the underlying dynamics of human driving behavior, which can be complex due to vehicle interactions and external factors, by learning from observed trajectories, e.g., sequences of positions, speeds, and accelerations. We begin by formulating the problem, motivating the need for modeling temporal correlations in driving behaviors, and then presenting a novel framework with both stationary and nonstationary kernels to explicitly model of the temporal correlations based on GPs.

\subsection{Problem Formulation}
Car-following models describe the interaction between a leading and a following vehicle, collectively referred to as a \textit{car-following pair}. These models aim to predict the acceleration of the following vehicle at time $t$ based on certain key covariates, including the inter-vehicle distance gap $s_{t}$, the speed of the following vehicle $v_{t}$, and the relative speed $\Delta v_{t}$ between the following and leading vehicles. These covariates reflect the information a driver uses to make acceleration decisions that balance safety, comfort, and traffic efficiency.

Let $t \in \{1, \dots, T\}$ index discrete time steps. We denote the sequential observations as $\{\boldsymbol{s}, \Delta\boldsymbol{v}, \boldsymbol{v}\}$, where $\boldsymbol{s} = [s_{1}, \dots, s_{T}] \in \setR^T$, $\Delta \boldsymbol{v} = [\Delta v_{1}, \dots, \Delta v_{T}] \in \setR^T$, and $\boldsymbol{v} = [v_{1}, \dots, v_{T}] \in \setR^T$. The goal is to predict the following vehicle's acceleration sequence $\boldsymbol{a} = [a_{1}, \dots, a_{T}] \in \setR^T$ given these inputs $\{{s}_t, \Delta{v}_t, {v}_t\}$. A general car-following model can be expressed as:
\begin{equation}
    {a}_t = f_{\mathrm{CF}}({s}_t, \Delta {v}_t, {v}_t; \boldsymbol{\theta}),
\end{equation}
where $f_{\mathrm{CF}}(\cdot)$  denotes the car-following function parameterized by $\boldsymbol{\theta}$. This function typically models highly nonlinear and dynamic interactions between covariates to predict the acceleration $\boldsymbol{a}$ of the following vehicle. 

Traditional car-following models, such as the IDM \citep{treiber2000congested}, represent $f_{\mathrm{CF}}(\cdot)$ analytically. These approaches simplify the complexity of driving behaviors by assuming static functional relationships between the covariates. However, human driving behaviors exhibit significant temporal variability and uncertainty, which deterministic models often fail to capture. In the following, we aim to enhance the modeling of car-following behaviors by leveraging sequential data $\{\boldsymbol{s}, \Delta\boldsymbol{v}, \boldsymbol{v}\}$ to account for the inherent stochasticity and temporal dependencies in driving behaviors.

\subsection{Motivations and Theoretical Background}
From a statistical perspective, modeling car-following behavior is essentially a regression task. 
Classical regression methods, such as ordinary least squares (OLS), assume homoskedastic and temporally uncorrelated errors, i.e., the residual covariance matrix $\boldsymbol{\Sigma}$ is proportional to the identity matrix, where errors exhibit constant variance and no temporal correlations. However, real-world human driving behaviors violate these assumptions due to 
\begin{enumerate}
    \item \textbf{Heteroskedasticity}, where uncertainty varies with the driving context, which evolves over time;
    \item \textbf{Temporal correlations}, where errors at adjacent time steps are statistically dependent \citep{zhang2024bayesian,zhang2024calibrating}.
\end{enumerate}
\textit{\textbf{Heteroskedasticity} leads to unequal diagonal elements of $\boldsymbol{\Sigma}$, reflecting context-dependent uncertainty levels, while \textbf{correlated errors} introduce non-zero off-diagonal elements, capturing temporal dependencies that vary dynamically with driving conditions.} Generalized least squares (GLS) extends OLS by incorporating a structured covariance matrix $\boldsymbol{\Sigma}$ that can handle these heteroskedastic and correlated errors.  This motivates our GP–based probabilistic framework, which explicitly models these temporal correlations and dynamic uncertainties using a nonstationary covariance structure adaptive to driving context, thereby enhancing both realism and robustness in car-following simulations.

With the insights from the above discussion, we adopt the formulation proposed in \citet{zhang2024bayesian} and \citet{zhang2024calibrating}, representing the action $a_t =a(\boldsymbol{x}_t, t)$ of a following vehicle at time $t$ as
\begin{equation}
    a_t = f_{\mathrm{CF}}(\boldsymbol{x}_t) + \delta(t) + \epsilon_t, \label{general_form}
\end{equation}
where $\boldsymbol{x}_t=[s_t,\Delta v_t, v_t]$ denotes the input covariates, $f_{\mathrm{CF}}(\boldsymbol{x}_t)$ as a function of $\boldsymbol{x}_t$ represents the mean car-following model, $\delta(t)$ accounts for temporal correlations, and $\epsilon_t \stackon{$\sim$}{\tiny{\textit{i.i.d.}}}\, \mathcal{N}(0,\sigma_0^2)$ is an \textit{independent and identically distributed (i.i.d.)} noise term with variance $\sigma_0^2$. {Note that $\delta(t)$ is the temporally correlated (colored-noise) component, which controls the correlation timescale and smoothness of residual accelerations. The additional term $\epsilon$ captures unstructured observation/estimation noise (e.g., differentiation-induced acceleration noise) and other high-frequency effects, improving identifiability between behavioral temporal dependence and measurement artifacts.}

In much of the prior literature, the term $\delta(t)$, representing temporally correlated components of driving behavior, is often omitted, which is an assumption that significantly oversimplifies the stochastic structure of car-following dynamics. This temporal component is, however, essential for capturing the intrinsic memory effects and autocorrelated decision-making processes exhibited by human drivers. Acceleration decisions are rarely made in isolation; instead, they are influenced by prior states, such as historical speeds, relative distances, and preceding actions. Disregarding $\delta(t)$ reduces the model to one in which all variability is attributed to the \textit{i.i.d.} noise term $\epsilon_t$, which leads to a misestimation of uncertainty and fails to reflect the dependent structure of the underlying process. Incorporating $\delta(t)$ allows for a more faithful representation of these dependencies, thereby enabling more realistic and interpretable simulations of car-following behavior.

A comparative overview of how the literature incorporates temporal correlation into car-following models is presented in \Cref{tab:literature}. \citet{treiber2006delays} introduced an OU process to model $\delta(t)$, which provides a continuous-time, mean-reverting formulation. However, the OU process is inherently linear and governed by a fixed decay rate, which imposes a rigid structure on temporal correlations and limits its ability to capture adaptive or nonlinear behaviors observed in real driving contexts. \citet{hoogendoorn2010calibration} eliminate $\delta(t)$ based on a Cochrane-Orcutt correction by assuming a first-order autoregressive (i.e., AR(1)) structure in the residuals. While computationally efficient, this approach also assumes a fixed correlation structure and linear dependence, constraining its capacity to model the complex, scenario-dependent dynamics that characterize real-world car-following behavior. More recent efforts, such as \citet{zhang2024bayesian}, used GPs with stationary kernels to capture smooth and consistent patterns over time. This approach improves flexibility by enabling smooth, nonparametric modeling of correlations. Nonetheless, the assumption of stationarity still imposes a uniform correlation structure over time, which is inadequate for modeling nonstationary phenomena such as sudden braking or transitions between driving regimes. In a related direction, \citet{zhang2024calibrating} adopted higher-order AR processes to capture both positive and negative temporal dependencies, offering improved expressiveness compared to AR(1) and OU processes, but remaining fundamentally limited by their stationary nature.

Although these approaches differ in formulation and modeling capacity, they are unified in their reliance on autocorrelation and mean-reversion as foundational principles. The AR(1) model, the OU process, and the Cochrane–Orcutt correction all assume exponentially decaying memory, differing mainly in whether the formulation is discrete-time, continuous-time, or embedded within a regression context. Extensions such as GPs and higher-order AR models offer increased flexibility but continue to operate under the assumption of stationary residual structures. This uniformity constrains their ability to adapt to temporally heterogeneous driving behavior, where reaction times, uncertainty tolerance, and behavioral regimes vary over time. Therefore, while effective in certain contexts, these methods fall short in modeling nonstationary dynamics, which are critical for capturing the full spectrum of human car-following behavior.

To provide a comprehensive understanding of our modeling approach, it is beneficial to distinguish the different forms of uncertainty encountered in traffic flow modeling. Broadly, uncertainty can be categorized into two primary groups \citep{zhou2025calibration}: (i) \textit{aleatoric uncertainty}, reflecting inherent randomness in driving behaviors such as variability in reaction times and driver aggressiveness; and (ii) \textit{epistemic uncertainty}, arising from limited knowledge, incomplete data, or model mis-specification, such as unobserved driver intentions or omitted explanatory variables. Within these broad categories, the primary focus of our approach, i.e., residual temporal correlation, is conceptually a structured form of \textit{aleatoric uncertainty}, capturing the memory effects and sequential dependencies in driver decisions not explained by the mean dynamics alone. However, it also overlap with \textit{epistemic uncertainty}, as persistent patterns in residuals may indicate missing explanatory variables or latent behavioral mechanisms. Our framework explicitly addresses this structured \textit{aleatoric uncertainty} through a nonstationary Gaussian process kernel, thus enhancing the realism and interpretability of stochastic simulations. Explicit modeling of \textit{epistemic uncertainty} remains a valuable direction for future research, potentially through hierarchical or latent variable approaches.


\begin{table}[t]
    \small
    \centering
    \caption{Modeling of temporal correlations in the literature, corresponds to \Cref{general_form}.}
    \begin{tabular}{p{3.4cm}lp{5.7cm}l}
    \toprule
    Reference & $f_{\mathrm{CF}}(\boldsymbol{x})$ & $\delta(t)$ & Nonstationary? \\
    \hline
    \citet{treiber2006delays}     &  IDM& Ornstein-Uhlenbeck (OU) processes & \xmark  \\
    \citet{hoogendoorn2010calibration} &  GHR/IDM& Cochrane-Orcutt correction \par (i.e., AR(1) process)& \xmark\\
    \citet{zhang2024bayesian} &  IDM & Gaussian processes & \xmark \\
    \citet{zhang2024calibrating} &  IDM & AR processes with higher orders & \xmark\\
    This work &  NN & Nonstationary GPs & \cmark \\
    \bottomrule
    \end{tabular}
    \label{tab:literature}
\end{table}

\subsection{Temporal Dependencies Modeling in Car-Following Behaviors}
Given the inherently complex and nonlinear dependencies between acceleration, speed, position, and various latent factors, conventional formula-based car-following models often struggle to capture the diversity of behaviors observed across a large driver population. These models typically rely on fixed analytical forms with limited expressiveness, which hinders their ability to adapt to nuanced, data-driven behavioral patterns. To overcome this limitation, we reformulate $f_{\mathrm{CF}}(\boldsymbol{x}_t)$ as a neural network function $a_\mathrm{NN}(\boldsymbol{h}_t; \boldsymbol{\theta})$, parameterized by learnable $\boldsymbol{\theta}$. The hidden state $\boldsymbol{h}_t=\mathcal{F}(\boldsymbol{h}_{t-1},\boldsymbol{x}_t;{\boldsymbol{\theta}})$ is recursively updated using a recurrent architecture such as an LSTM or GRU, where $\mathcal{F}$ denotes the transition dynamics of the chosen RNN cell. This hidden state serves as a temporal memory that captures the influence of past covariates $\boldsymbol{x}_t$ on current acceleration decisions, thereby enabling the model to incorporate behavioral history into its predictions.

To explicitly model the structured temporal dependencies beyond those captured by the RNN, it is crucial to account for the inherent nonlinear dependencies in the data, both within and across temporal observations. Here we adopt a Gaussian processes-based approach (see \citet{zhang2024bayesian} for more details) to capture these dependencies, which provides a richer and more realistic representation than a simple identity covariance matrix from the inappropriate \textit{i.i.d.} assumption, yielding the hybrid formulation:
\begin{equation}\label{eq3}
    a_t = a_\mathrm{NN}(\boldsymbol{h}_t;\boldsymbol{\theta}) + \delta_\mathrm{GP}(t; \boldsymbol{\lambda}) + \epsilon_t,
\end{equation}
where $\delta_\mathrm{GP}(t; \boldsymbol{\lambda})\sim \mathcal{GP}(0, k(t,t'; \boldsymbol{\lambda}))$ is a zero-mean GP governed by a kernel function $k$, parameterized by $\boldsymbol{\lambda}$. {The additional term $\epsilon$ is reserved for unstructured, high-frequency effects, effectively acting as a ``nugget effect''\footnote{The nugget effect refers to the variance of the i.i.d. noise term $\epsilon$, capturing measurement errors and micro-scale variability occurring at scales smaller than the sampling frequency. This ensures that the GP kernel prioritizes structured behavioral dependencies rather than high-frequency sensor noise.} commonly used in GP modeling.}  This formulation augments the neural model with a temporally correlated residual process, effectively decoupling the modeling of nonlinear mean dynamics from that of structured uncertainty. Over a finite time horizon of $T$ steps with sampling interval $\Delta t$, the GP prior over the sequence of temporal residuals $\boldsymbol{\delta}_\mathrm{GP}=\left[\delta_\mathrm{GP}(\Delta t),\ldots,\delta_\mathrm{GP}(T\Delta t)\right]^{\top}$ is given by a zero-mean Gaussian distribution $\boldsymbol{\delta}_\mathrm{GP}\sim\mathcal{N}(\boldsymbol{0},\boldsymbol{K})$, where $\boldsymbol{K}\in\setR^{T\times T}$ is the kernel matrix (also known as Gram matrix) with entries $[\boldsymbol{K}]_{ij}=k(i\Delta t,j\Delta t;\boldsymbol{\lambda})$.

Inspired by the ``better batch'' training strategy proposed by \citet{zheng2024better}, we consider \Cref{eq3} in a vector form over a mini-batch of $\Delta T$ consecutive time steps. The model's likelihood over the mini-batch can then be written compactly as:
\begin{align}\label{llh_general}
    \underbrace{\left[\begin{array}{c}
         a_{1}  \\
         \vdots \\
         a_{\Delta T}  \\ 
    \end{array}\right]}_{\boldsymbol{a}^{\text{batch}}}
     \sim \mathcal{N}\Bigg(\underbrace{\left[\begin{array}{c}
        a_\mathrm{NN}(\boldsymbol{h}_1; \boldsymbol{\theta}) \\
        \vdots\\
        a_\mathrm{NN}(\boldsymbol{h}_{\Delta T}; \boldsymbol{\theta}) \\
    \end{array}\right]}_{\boldsymbol{a}^{\text{batch}}_{\text{NN}}}, \underbrace{\left[ \begin{array}{ccc}
        k(\Delta t, \Delta t; \boldsymbol{\lambda}) &  \cdots & k(\Delta t, \Delta T \Delta t; \boldsymbol{\lambda}) \\
        \vdots & \ddots & \vdots \\
        k(\Delta T \Delta t, \Delta t; \boldsymbol{\lambda}) & \cdots & k(\Delta T \Delta t, \Delta T \Delta t; \boldsymbol{\lambda})
    \end{array} \right]}_{\boldsymbol{K}} + \sigma_0^2\boldsymbol{I}_{\Delta T}\Bigg),
\end{align}
where $\boldsymbol{I}_{\Delta T}$ denotes a $\Delta T\times \Delta T$ identity matrix, and $\sigma_0^2 \boldsymbol{I}_{\Delta T}$ captures homoskedastic observation noise.

This kernel-based formulation offers a significant advantage over the common \textit{i.i.d.} assumption. By encoding structured correlations between time steps, the kernel matrix $\boldsymbol{K}$ reflects the intuition that temporally adjacent data points are more strongly related. This enables the model to better capture behavioral patterns such as gradual acceleration, emergency braking, or consistent following distances. Furthermore, the structured covariance accounts for temporally correlated noise sources such as sensor drift or lane tracking errors, yielding more informative uncertainty quantification. From a simulation perspective, this approach enables tighter confidence bounds in periods of stable driving, while naturally widening uncertainty intervals during erratic behavior or transitions between regimes. It also promotes temporal smoothness in the predicted acceleration trajectory, which is essential for maintaining physical plausibility and ensuring the continuity of vehicle dynamics.

\subsection{Kernel Design for Capturing Temporal Dependencies}
In our probabilistic framework, temporal dependencies in driver behavior are encoded via a kernel function $k(t,t';\boldsymbol{\lambda})$ that defines the covariance structure of the GP residual $\delta_\mathrm{GP}(t;\boldsymbol{\lambda})$. This section introduces three kernel choices: two stationary kernels (squared exponential and Mat\'{e}rn) and one nonstationary kernel (Gibbs). These kernels differ in their ability to capture temporal smoothness, adaptability, and heteroskedasticity, which are essential for modeling realistic car-following behaviors.

\subsubsection{Stationary kernel: Squared exponential (SE)}
The squared exponential (SE) kernel, also known as the Radial Basis Function (RBF) kernel, is one of the most commonly used kernels in GP modeling. It assumes smooth and stationary correlations between time steps, defined as
\begin{equation}
    k_{\text{SE}}(t,t';\boldsymbol{\lambda}) := \sigma^2\,\mathrm{exp}\left( -\frac{d(t, t')^2}{2 \ell^2} \right),
\end{equation}
where $\boldsymbol{\lambda}=[\ell,\sigma]$ includes the lengthscale $\ell$ and the marginal variance $\sigma^2$, and $d(t,t')=\parallel t-t'\parallel_2$ denotes the Euclidean distance between time steps. The lengthscale $\ell$ controls the decay rate of temporal correlations, shorter values result in rapidly decaying correlations.

The SE kernel assumes that the latent function is infinitely differentiable, making it suitable for modeling smooth, continuous processes. {However, this smoothness assumption can be restrictive in systems exhibiting \emph{rapid regime transitions} (e.g., short corrective braking/acceleration bursts), where the effective correlation timescale changes markedly even though acceleration remains physically continuous.
} In such cases, the SE kernel can lead to over-smoothed predictions that fail to capture rapid behavioral changes.

\subsubsection{Stationary kernel: Mat\'{e}rn}
To address the limitations of the SE kernel, the Mat\'{e}rn kernel provides a more flexible framework for controlling the smoothness of the latent function. It is defined as
\begin{equation}
    k_{\text{Mat\'{e}rn}}^{\nu}(t, t';\boldsymbol{\lambda}):=\sigma ^{2}{\frac {2^{1-\nu }}{\Gamma (\nu)}}{\Bigg (}{\sqrt {2\nu }}{\frac {d(t, t')}{\ell }}{\Bigg )}^{\nu }K_{\nu }{\Bigg (}{\sqrt {2\nu }}{\frac {d(t, t')}{\ell }}{\Bigg )},
\end{equation}
where $\nu>0$ is a smoothness parameter, $\Gamma(\cdot)$ is the gamma function, $K_{\nu}(\cdot)$ is the modified Bessel function of the second kind. It is worth noting that the Mat\'{e}rn kernel is $\lfloor\nu\rfloor$-times differentiable, offering a tunable tradeoff between smoothness and expressiveness. For instance, smaller $\nu$ leads to rougher functions with long-range correlations, while larger $\nu$ approaches the behavior of the SE kernel.

A commonly used variant for modeling traffic-related time series is the Mat\'{e}rn kernel with $\nu=5/2$, which balances smoothness and flexibility, written as
\begin{equation}
    k_{\text{Mat\'{e}rn}}^{5/2}(t, t';\boldsymbol{\lambda}):=\sigma ^{2}\left(1+{\frac {{\sqrt {5}}d(t, t')}{\ell}}+{\frac {5d(t, t')^{2}}{3\ell ^{2}}}\right)\exp \left(-{\frac {{\sqrt {5}}d(t, t')}{\ell}}\right).
\end{equation}
This formulation is particularly suitable for modeling human driving behaviors, which often involve irregular and non-smooth transitions.

\subsubsection{Nonstationary kernel: Gibbs}
While stationary kernels assume that temporal correlations are invariant over time, real-world driving behavior often exhibits nonstationary dynamics, drivers’ reactions and uncertainties vary across different scenarios. To model such behavior, we adopt the Gibbs kernel, a nonstationary generalization of the SE kernel introduced by \citet{gibbs1997bayesian}. In this formulation, both the variance $\sigma^2(t)$ and the lengthscale $\ell(t)$ are treated as input-dependent functions:
\begin{equation}
    k_{\text{Gibbs}}(t, t'; \boldsymbol{\lambda}) := \sigma(t) \sigma(t') \underbrace{\sqrt{\frac{2 \ell(t) \ell(t')}{\ell(t)^2 + \ell(t')^2}} \exp{\left(-\frac{(t - t')^2}{\ell(t)^2 + \ell(t')^2}\right)}}_{k_{\text{Gibbs}}^{\star}(t, t'; \boldsymbol{\lambda})},
\end{equation}
where $\boldsymbol{\lambda}(t)=[\ell(t), \sigma(t)]$ are parameterized by a neural network conditioned on contextual inputs. This kernel effectively captures both the context-dependent smoothness and heteroskedasticity in the residual process. Following \citet{zheng2024better}, an efficient implementation of the Gibbs kernel over a mini-batch of time steps $t\in\{\Delta t, 2\Delta t, \dots,\Delta T\Delta t\}$ can be constructed as
\begin{equation}
    \boldsymbol{K}=
    \text{diag}(\boldsymbol{\sigma}^{\text{batch}})\boldsymbol{K}^{\star}\text{diag}(\boldsymbol{\sigma}^{\text{batch}}), \label{nonstationary_matrix_form}
\end{equation}
where $\boldsymbol{\sigma}^{\text{batch}}=[\sigma(\Delta t), \dots,\sigma(\Delta T\Delta t)]$ and $\boldsymbol{K}^{\star}=[k_{\text{Gibbs}}^{\star}(t, t'; \boldsymbol{\lambda})]$ is the base kernel matrix prior to applying variance modulation. This factorization enables scalable computation of nonstationary kernels during training.

Notably, the Gibbs kernel simultaneously captures two critical properties: \textbf{Heteroskedasticity} via context-dependent variance $\sigma^2(\boldsymbol{x}_t)$, representing time-varying uncertainty; and \textbf{Nonstationary correlations} via context-adaptive lengthscale $\ell(t)$, encoding dynamic adaptation in temporal dependence. These capabilities make the Gibbs kernel especially well-suited for modeling stochastic driving behaviors that are both context-dependent and temporally heterogeneous.

\subsection{Neural Network Parameterization for Probabilistic Modeling}
To capture both the mean dynamics and temporal uncertainty structure of car-following behavior, we propose parameterizing the key components of the model, i.e., the mean acceleration, kernel lengthscale, and variance, using neural networks. { In this paper we use DeepAR \citep{salinas2020deepar} to instantiate the mean model $f_{\mathrm{CF}}(\cdot)$, but the stochastic residual component $\delta(t)$ is model-agnostic and can be combined with alternative car-following formulations.}

DeepAR is an RNN architecture designed for time series forecasting, wherein the temporal dependencies among observations are captured via a hidden state $\boldsymbol{h}_t$, which evolves over time as
$\boldsymbol{h}_t = \mathcal{F}(\boldsymbol{h}_{t-1}, \boldsymbol{x}_t; \boldsymbol{\theta})$, where $\mathcal{F}$ denotes the recurrent dynamics (e.g., LSTM or GRU), and $\boldsymbol{\theta}$ represents the model parameters. The hidden state $\boldsymbol{h}_t$ at time $t$ summarizes both historical and current input information. At each time step $t$, we use the hidden state to produce three outputs (mean acceleration, kernel lengthscale, and variance) through separate nonlinear transformations\footnote{\textbf{Clarification on ``time-varying'' vs.\ ``context-dependent''}: It is important to clarify the distinction between ``time-varying'' and ``context-dependent'' in our model. While the outputs---mean acceleration $a_{\mathrm{NN}}$, kernel lengthscale $\ell_{\mathrm{NN}}$, and variance $\sigma_{\mathrm{NN}}$---are indexed by discrete time steps $t$, their values are not directly functions of time. Instead, they are derived from the hidden state $\boldsymbol{h}_t$, which summarizes the evolving input context $\boldsymbol{\mathcal{X}}_t$ (e.g., spacing, speed, relative speed). As such, these parameters are \textit{context-dependent}---their values depend on the driving state and its temporal history, not on time itself. Because the context evolves over time, the resulting behavior appears time-varying, but the underlying dependency is mediated by dynamic driving conditions. This distinction emphasizes that the model adapts to changing scenarios, rather than assuming any explicit temporal structure a priori.}:
\begin{subequations}
\begin{align}
    a_\mathrm{NN}(\boldsymbol{h}_t) &= \mathcal{F}_{\mu}(\boldsymbol{h}_t)\in \setR, \\
    \ell_\mathrm{NN}(\boldsymbol{h}_t) &= \log\left(1 + \exp\left(\mathcal{F}_{\ell}(\boldsymbol{h}_t)\right)\right)\in \setR, \\
    \sigma_\mathrm{NN}(\boldsymbol{h}_t) &= \log\left(1 + \exp\left(\mathcal{F}_{\sigma}(\boldsymbol{h}_t)\right)\right)\in \setR,
\end{align}
\end{subequations}
where $\mathcal{F}_{\mu}$, $\mathcal{F}_{\ell}$, and $\mathcal{F}_{\sigma}$ are feedforward (fully connected) layers applied to the hidden state $\boldsymbol{h}_t$. The softplus transformation ensures positivity of the predicted lengthscale and variance.

Let $T$ denote the context length. Given a time series input $\boldsymbol{\mathcal{X}}_t=[\boldsymbol{x}_{t-T+1},\dots,\boldsymbol{x}_{t}]\in\setR^{3\times T}$, we compute the outputs at time $t$, i.e., $a_\mathrm{NN}(\boldsymbol{h}_t)$, $\ell_\mathrm{NN}(\boldsymbol{h}_t)$, and $\sigma_\mathrm{NN}(\boldsymbol{h}_t)$. For a mini-batch containing $\Delta T$ such time series segments, $\boldsymbol{\mathcal{X}}^{\text{batch}}= \{\boldsymbol{\mathcal{X}}_{t}, \dots, \boldsymbol{\mathcal{X}}_{t+\Delta T}\}$ with $\Delta T$ as the batch size, the model produces
\begin{align}\nonumber
    \boldsymbol{a}_{\text{NN}}^{\text{batch}}&=[a_{\mathrm{NN}}( \boldsymbol{h}_{t}), \dots, a_{\mathrm{NN}}( \boldsymbol{h}_{t+\Delta T})]\in\setR^{\Delta T},\\\nonumber
    \boldsymbol{\ell}_{\text{NN}}^{\text{batch}} &=[\ell_{\mathrm{NN}}( \boldsymbol{h}_{t}), \dots, \ell_{\mathrm{NN}}( \boldsymbol{h}_{t+\Delta T})]\in\setR^{\Delta T},\\\nonumber
    \boldsymbol{\sigma}_{\text{NN}}^{\text{batch}} &=[\sigma_{\mathrm{NN}}( \boldsymbol{h}_{1}), \dots, \sigma_{\mathrm{NN}}( \boldsymbol{h}_{t+\Delta T})]\in\setR^{\Delta T}.
\end{align}

To construct the kernel matrix $\boldsymbol{K}$, we proceed as follows: for stationary kernels (e.g., SE or Mat\'{e}rn), we use the batch-averaged parameters $\bar\ell_{\mathrm{NN}}=\mathrm{mean}(\boldsymbol{\ell}_{\text{NN}}^{\text{batch}})$ and $\bar\sigma_{\mathrm{NN}}=\mathrm{mean}(\boldsymbol{\sigma}_{\text{NN}}^{\text{batch}})$; for the nonstationary Gibbs kernels, the full sequences $\boldsymbol{\ell}_{\text{NN}}^{\text{batch}}$ and $\boldsymbol{\sigma}_{\text{NN}}^{\text{batch}}$ are used to construct $\boldsymbol{K}$ via the formulation in \Cref{nonstationary_matrix_form}.

Given the neural predictions, the conditional likelihood of the observed acceleration sequence $\boldsymbol{a}^{\text{batch}}$ follows the multivariate Gaussian
\begin{equation}\label{a_likelihood}
    \boldsymbol{a}^{\text{batch}}|\boldsymbol{a}_{\text{NN}}^{\text{batch}},\boldsymbol{\ell}_{\text{NN}}^{\text{batch}},\boldsymbol{\sigma}_{\text{NN}}^{\text{batch}},\boldsymbol{\theta}\sim\mathcal{N}(\boldsymbol{a}_{\text{NN}}^{\text{batch}}, \boldsymbol{K}+\sigma_0^2 \boldsymbol{I}_{\Delta T}),
\end{equation}
where $\sigma_0^2 \boldsymbol{I}_{\Delta T}$ models homoskedastic observation noise.  This formulation allows the model to output both predictions and temporally structured uncertainty, modulated by the GP kernel. The neural network is trained by minimizing the negative log-likelihood of \Cref{a_likelihood}, with an added regularization term $\mathcal{L}_{\text{reg}}$ to prevent overfitting
\begin{equation}\label{opt_problem}
    \boldsymbol{\theta}^{\star}=\argmin_{\boldsymbol{\theta}}\, \frac{1}{2} \log |\boldsymbol{K}+\sigma_0^2 \boldsymbol{I}_{\Delta T}| + 
\frac{1}{2} (\hat{\boldsymbol{a}}^{\text{batch}} - \boldsymbol{a}_{\text{NN}}^{\text{batch}})^\top 
(\boldsymbol{K}+\sigma_0^2 \boldsymbol{I}_{\Delta T})^{-1} 
(\hat{\boldsymbol{a}}^{\text{batch}} - \boldsymbol{a}_{\text{NN}}^{\text{batch}}) + \mathcal{L}_{\text{reg}}.
\end{equation}
Here $\hat{\boldsymbol{a}}^{\text{batch}}$ denotes the observed accelerations, and $\boldsymbol{a}^{\text{batch}}_{\mathrm{NN}}$ are the corresponding model predictions. The constant term $n\log(2\pi)/2$ is omitted as it does not affect the optimization.

This end-to-end probabilistic framework enables the network to learn not only a data-driven approximation of the car-following dynamics, but also an expressive, context-dependent model of uncertainty, enhancing both interpretability and simulation fidelity. 

\section{Stochastic Simulations and Empirical Evaluations}\label{sec:exp_and_results}

\begin{figure}[!t]
    \centering
    \includegraphics[width=.97\linewidth]{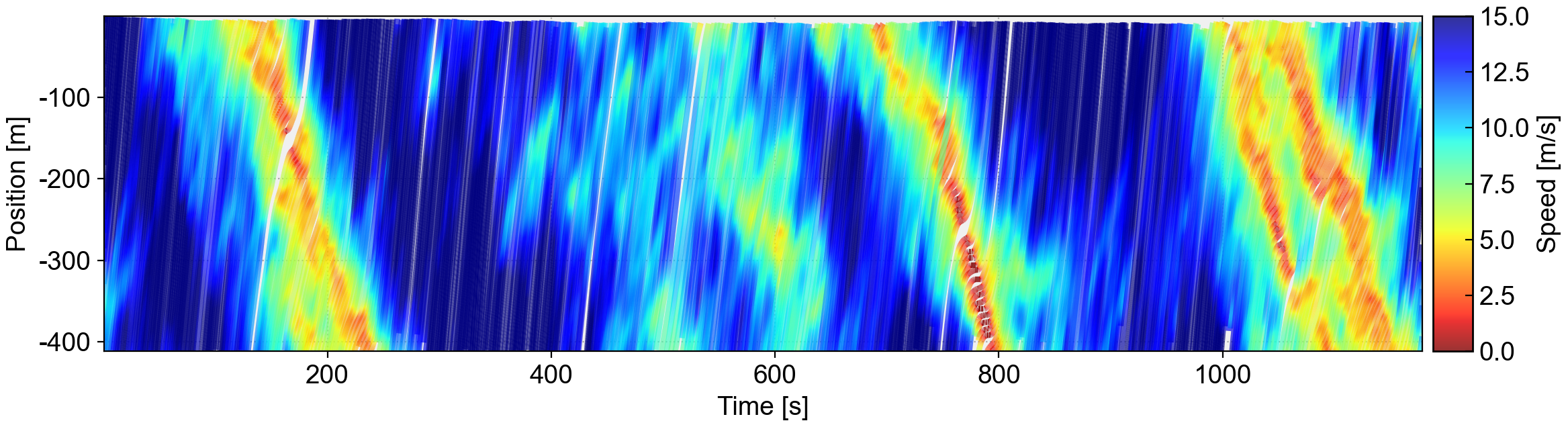}
    \vspace{-10pt}
    \caption{Traffic shockwave visualization of HighD dataset (Track \#25, Lane \#4), colored by speed (m/s).}
    \label{fig:highD}
\end{figure}

\subsection{Dataset and Experimental Setup}
To evaluate the performance of the proposed DeepAR-based car-following model, we conduct experiments on the HighD dataset \citep{highDdataset}, which provides high-resolution vehicle trajectory data captured by drones over German highways. Compared to the NGSIM dataset \citep{punzo2011assessment}, HighD offers improved reliability due to enhanced data collection and post-processing techniques. The dataset consists of 60 drone-recorded videos, each covering a 420-meter stretch of highway, with detailed annotations for vehicle trajectories, velocities, and accelerations. {Here we choose HighD because acceleration-based stochastic modeling is sensitive to measurement noise, and HighD provides high-quality trajectories with widely adopted preprocessing conventions. While HighD is a freeway dataset, it contains multiple traffic states, including close-following and stop-and-go episodes (as shown in \Cref{fig:highD}), which are simple but sufficient to evaluate regime-dependent temporal correlation.
}

To reduce redundancy and enhance computational efficiency, we downsample the original data to $5$ Hz (i.e., $\Delta t=0.2$ s) by uniformly selecting every fifth frame. Following the preprocessing pipeline described in \citet{zhang2021spatiotemporal}, we transform the trajectory coordinates to a local vehicle-centric frame.

{To emphasize challenging interactions within HighD, we include sustained close-following episodes ($t_0 > 30$\,s) and evaluate performance across a range of traffic states, including stop-and-go behavior.} The resulting dataset is denoted as $\mathcal{D}_{\text{HighD}}$. From this, we extract a test set  $\mathcal{D}_{\text{test}}$ comprising 30 randomly chosen pairs with even longer durations ($t_0>50$) for qualitative evaluation and visualization. The remaining pairs are split into training and validation sets: $\mathcal{D}_{\text{train}}$ contains $70\%$ of the data, while $\mathcal{D}_{\text{val}}$ holds the remaining $30\%$.

For training the DeepAR model, we employ a one-layer LSTM with 64 hidden units. A linear transformation is applied to the final hidden state to predict distributional parameters. A dropout rate of 0.1 is used to mitigate overfitting. The model is optimized using Adam with a learning rate of $1\times10^{-3}$, a batch size of 256, and trained for up to 500 epochs. Regularization includes weight decay ($1\times10^{-3}$) and gradient clipping (maximum norm of 1) to ensure numerical stability. The Sigmoid Linear Unit (SiLU) activation function \citep{hendrycks2016gaussian} is employed in the feedforward layers.

For temporal correlation modeling, we primarily adopt the nonstationary Gibbs kernel. For comparison, we also evaluate variants based on the squared exponential (SE) kernel, Mat\'{e}rn kernel, and a white noise kernel  (i.e., using $\sigma^2\boldsymbol{I}_{\Delta T}$ instead of $\boldsymbol{K}$ as the covariance matrix in \Cref{a_likelihood} based on the \textit{i.i.d.} assumption).

\begin{figure}[!t]
    \centering
    \includegraphics[width=.8\linewidth]{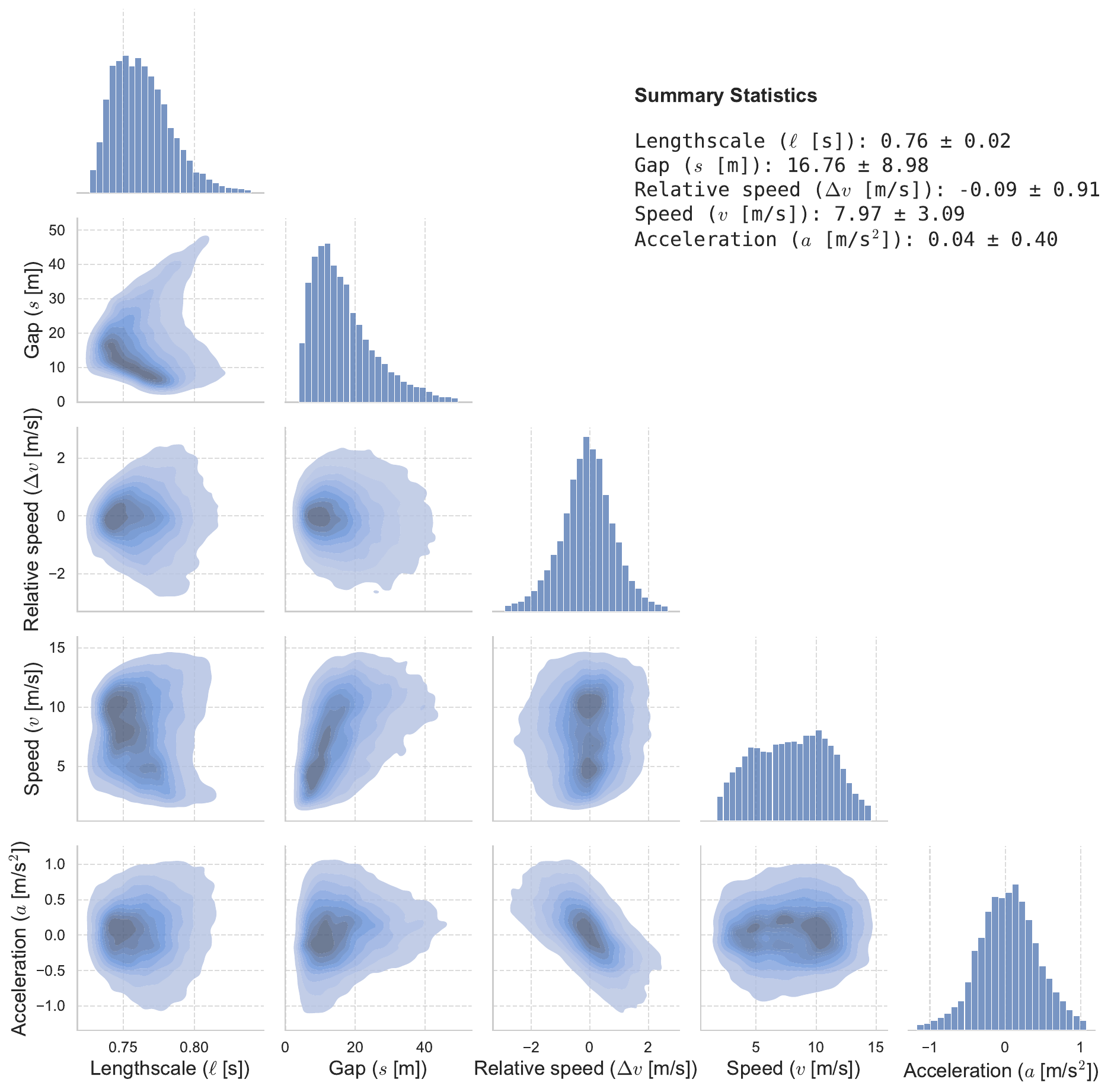}
    \caption{Joint and marginal distributions of the training‐set covariates along with the learned lengthscale. On the diagonal, the histograms show the empirical distribution of each variable: lengthscale $\ell$, gap $s$, relative speed $\Delta v$, speed $v$, and acceleration $a$. The off‐diagonal panels display bivariate contour estimates (kernel density plots) for each pair of variables, illustrating correlations between $\ell$ and the four car‐following covariates. In the top‐right corner, summary statistics are provided for reference.}
    \label{fig:Data_Density_Pairwise}
\end{figure}

\Cref{fig:Data_Density_Pairwise} visualizes the statistical properties of the training data.
The diagonal subplots show marginal histograms for the covariates gap $s$, relative speed $\Delta v$, speed $v$, and acceleration $a$, alongside the learned lengthscale $\ell$. Notably, $\ell$ exhibits a sharp peak around 0.76 s, indicating that under typical conditions, the learned model assumes a short temporal memory for car-following behavior. In contrast, the behavioral covariates exhibit broader and more varied distributions: spacing ranges from 0 to over 50 meters, speeds span both stop-and-go and free-flow regimes, and accelerations cluster around zero but with visible tails. The bimodality in the speed histogram suggests the presence of multiple traffic states, such as congestion and cruising.

The joint distributions offer further insight into the adaptive behavior learned by the model. For instance, sharp braking behavior (characterized by large negative accelerations) coincides with small gaps and high positive relative speeds, confirming expected physical driving responses. Besides, the learned lengthscale $\ell$ and spacing $s$ reveal a non-monotonic relationship that changes direction around a critical threshold of approximately 20 meters. For small gaps ($s\lesssim 20$ m), a weak negative correlation emerges, where $\ell$ tends to decrease as the gap increases. In very tight headways ($s<10$ m), $\ell$ is slightly elevated---reflecting smoother, more cautious behavior under critical proximity. As spacing grows toward 20 m, drivers become more reactive, resulting in shorter memory and a contraction of $\ell$; for large gaps ($s\gtrsim 20$ m), the trend reverses---a positive correlation appears, with $\ell$ gradually increasing alongside $s$. In these wider spacings, drivers adopt more stable and anticipatory control strategies, prompting the model to expand its temporal correlation horizon. These findings suggest that the learned kernel dynamically adapts its memory scale based on the perceived driving context. By presenting both marginals and joint densities, this figure emphasizes how $\ell$ co‐varies with traffic states and highlights the model’s ability to learn a concise, context‐dependent correlation horizon from the training data.

\subsection{Multi-Round Stochastic Simulations}
\subsubsection{Simulation formulation}
To emulate realistic car-following behavior under uncertainty, we adopt a stochastic simulation framework based on GP with RNN dynamics. This formulation allows us to capture both the mean trajectory and temporally correlated variability in acceleration decisions, producing simulation rollouts that reflect the inherent randomness of human driving. Further technical background on this formulation is available in \citet{zhang2024bayesian, zhang2024calibrating}.

At each time step $T+1$, the predictive distribution of acceleration is obtained by conditioning the GP on the observed sequence of past accelerations $\hat{\boldsymbol{a}}_{1:T}$ and corresponding inputs $\boldsymbol{x}_{1:T+1}$. Specifically, the predicted acceleration follows a univariate Gaussian distribution:
\begin{align}\label{sample_acc}
    a_{T+1} | \hat{\boldsymbol{a}}_{1:T}; \boldsymbol{x}_{1:T+1};\boldsymbol{\theta} \sim \mathcal{N}(\mu_{T+1}, \sigma_{T+1}^2),
\end{align}
with the conditional mean $\mu_{T+1}$ and variance $\sigma_{T+1}^2$ computed as:
\begin{subequations}\label{sample_mu_and_sigma}
\begin{align}
    \mu_{T+1} &= a_{\mathrm{NN}}(\boldsymbol{h}_{T+1}; \boldsymbol{\theta}) + \boldsymbol{K}_* (\boldsymbol{K} + \sigma_0^2 \boldsymbol{I}_{T})^{-1} (\hat{\boldsymbol{a}}_{1:T} - \boldsymbol{a}_{\text{NN}}^{(1:T)}), \\
    \sigma_{T+1}^2 &= K_{**} - \boldsymbol{K}_* (\boldsymbol{K} + \sigma_0^2 \boldsymbol{I}_{T})^{-1} \boldsymbol{K}_*^\top.
\end{align}
\end{subequations}
Here, $a_{\mathrm{NN}}(\boldsymbol{h}_{T+1}; \boldsymbol{\theta})$ denotes the mean acceleration predicted by the DeepAR network at time $T+1$, and $\boldsymbol{a}_{\text{NN}}^{(1:T)}=[a_{\mathrm{NN}}( \boldsymbol{h}_{1}), \dots, a_{\mathrm{NN}}( \boldsymbol{h}_{T})]^\top$ is the corresponding sequence of predicted means for the conditioning inputs. The kernel matrix $\boldsymbol{K}$ is computed from the observed inputs $\boldsymbol{X}_{1:T}$, reflecting temporal correlations; $\boldsymbol{K}_*$ encodes the cross-covariance between past and current inputs; $K_{**}$ denotes the prior variance at $T+1$. The term $\sigma_0^2 \boldsymbol{I}_{\Delta T}$ accounts for observation noise present in the training data.

\begin{algorithm}[t]
\caption{A round of stochastic simulations of car-following behaviors based on rolling predictions.}
\begin{algorithmic}[1]\label{sim_1_round}
\STATE \textbf{Input:} $t_{\text{start}}$, $t_{\text{end}}$, $\{v^{\text{lead}}_t\}_{t=t_{\text{pre}}}^{t_{\text{end}}}$, $\{p^{\text{lead}}_t\}_{t=t_{\text{pre}}}^{t_{\text{end}}}$, $\{a_t\}_{t=t_{\text{pre}}}^{t_{\text{start-1}}}$, $\{v_t\}_{t=t_{\text{pre}}}^{t_{\text{start}}}$, and $\{p_t\}_{t=t_{\text{pre}}}^{t_{\text{start}}}$. \STATE \textbf{Initialize:} $v^{\text{sim}}_t = v_{t_{\text{start}}}$ and $p^{\text{sim}}_t = p_{t_{\text{start}}}$.
\FOR{$t = t_{\text{start}}$ \textbf{to} $t_{\text{end}}$}
    \STATE Sample the acceleration based on \Cref{sample_acc}: ${a}^{\text{sim}}_{t}|p^{\text{sim}}_{t},v^{\text{sim}}_{t},- \sim \mathcal{N}(\mu_{t}, \sigma_{t}^2)$;
    \STATE Update the speed: $v^{\text{sim}}_{t+1} = v^{\text{sim}}_t + a^{\text{sim}}_t \Delta t$; 
    \STATE Update the position: $p^{\text{sim}}_{t+1} = p^{\text{sim}}_t + \frac{1}{2}(v^{\text{sim}}_t + v^{\text{sim}}_{t+1})\Delta t$;
\ENDFOR
\STATE \textbf{Output:} Simulated accelerations $\{a^{\text{sim}}_t\}_{t=t_{\text{start}}}^{t_{\text{end}}}$, speeds $\{v^{\text{sim}}_t\}_{t=t_{\text{start}}}^{t_{\text{end}}}$, and positions $\{p^{\text{sim}}_t\}_{t=t_{\text{start}}}^{t_{\text{end}}}$.
\end{algorithmic}
\end{algorithm}

As outlined in \Cref{sim_1_round}, the simulation procedure operates in a rolling fashion: at each time step $t$, the model samples an acceleration $a^{\text{sim}}$ from the predictive distribution in \Cref{sample_acc}, which is then used to update the follower’s speed and position. Iteratively applying this process over the simulation horizon yields a single stochastic realization of the vehicle trajectory. This approach stands in contrast to conventional deterministic simulations, which propagate only point estimates. By explicitly modeling predictive uncertainty and temporal correlations, the proposed method produces a distribution over future trajectories that reflects both the learned dynamics and the stochasticity of human behavior. As a result, it provides a principled way to generate ensemble simulations that are crucial for probabilistic safety analysis, risk assessment, and behaviorally realistic traffic modeling.

\begin{figure*}[t]
    \centering
    \subfigure[Speed, gap, and relative speed trajectories of a 100-vehicle platoon.]{
        \centering
        \includegraphics[width=0.655\textwidth]{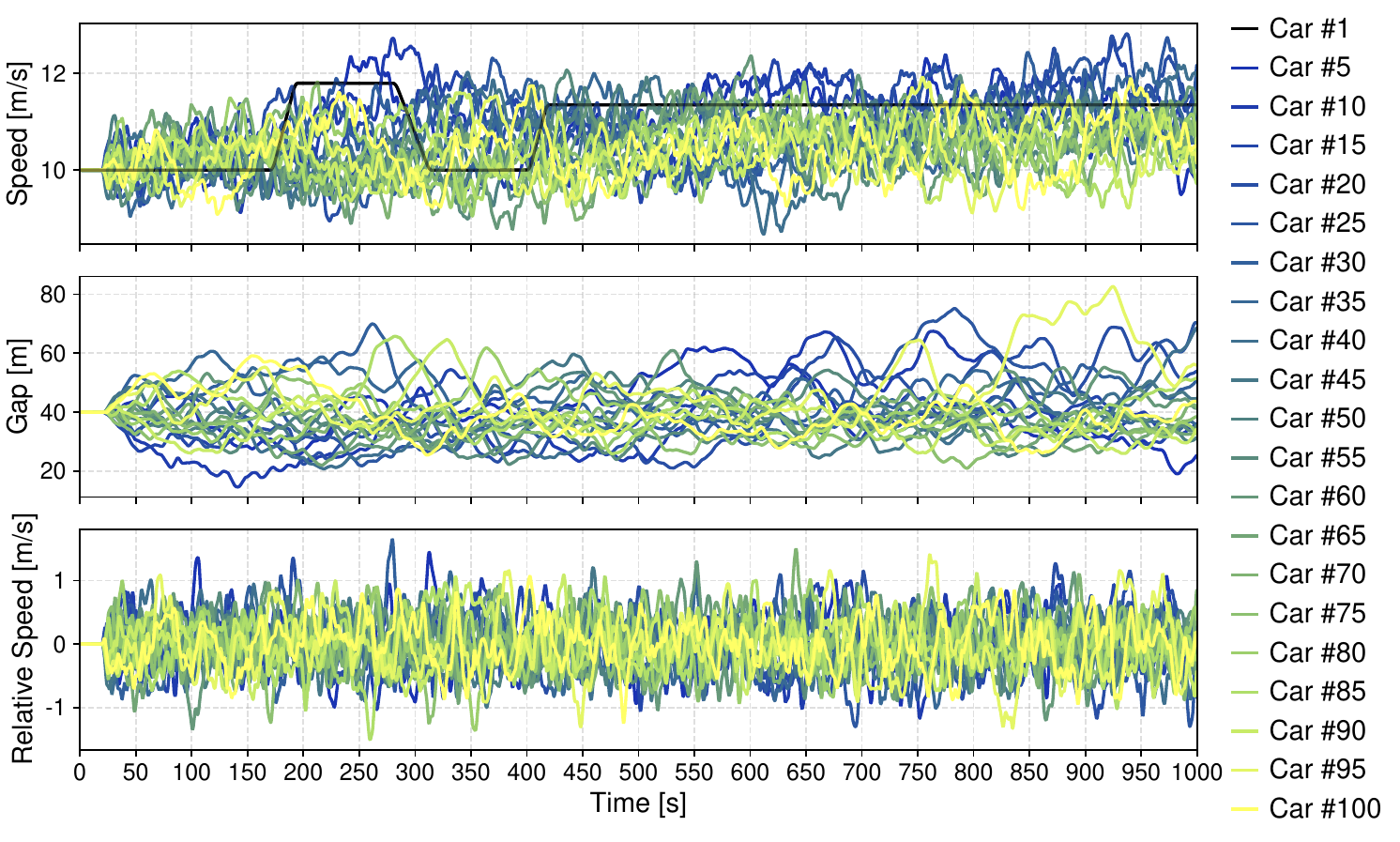}\label{platoon1}
    }
    \subfigure[Time–space diagram.]{
        \centering
        \includegraphics[width=0.315\textwidth]{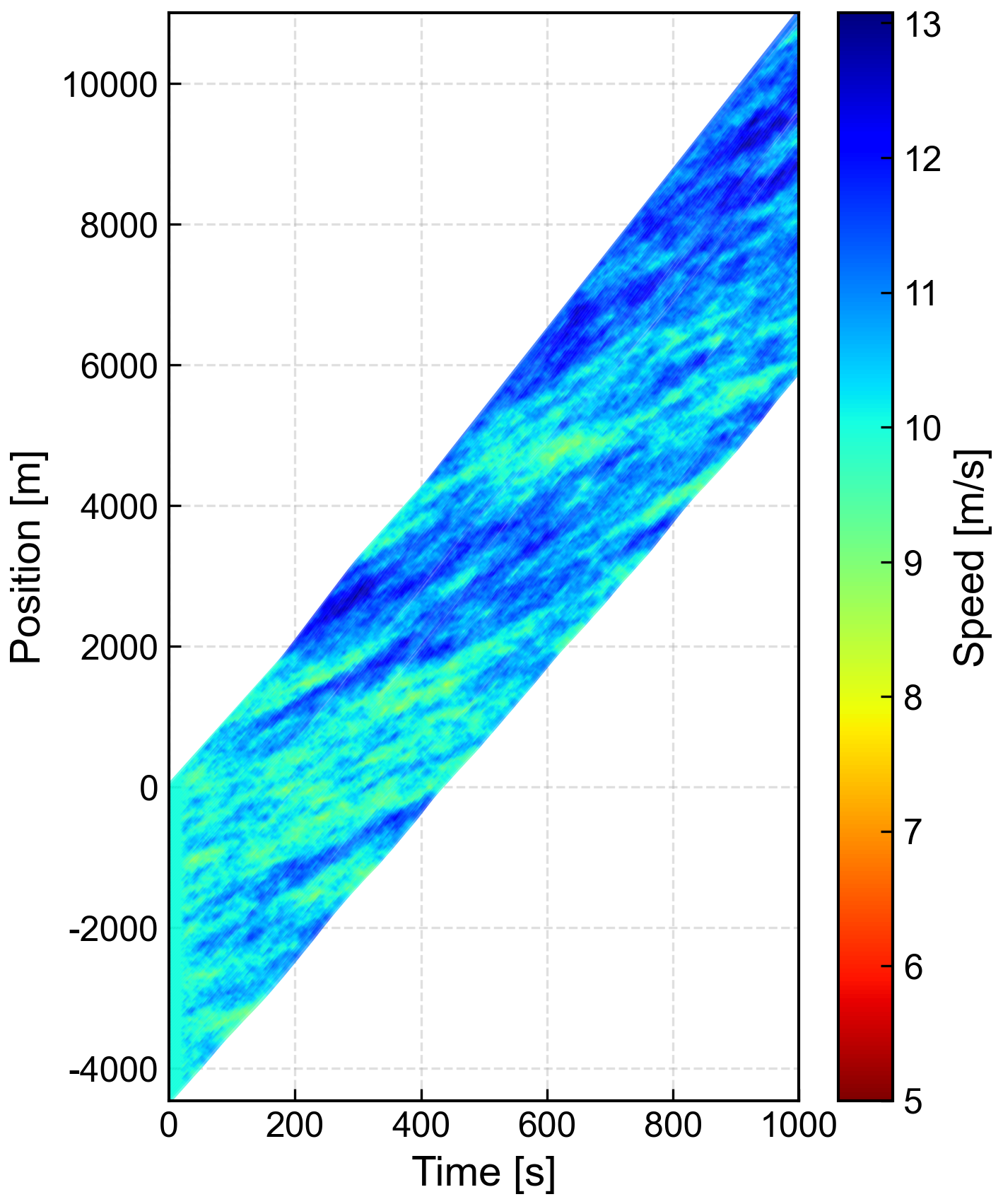}\label{time-space}
    }
    \caption{Simulated response of a 100-vehicle platoon to a smooth trapezoidal maneuver by the lead vehicle.}
    \label{fig:platoon}
\end{figure*}

\Cref{fig:platoon} provides a detailed visualization of the car-following dynamics in a 100-vehicle platoon simulation using \Cref{sim_1_round}. \Cref{platoon1} shows the individual trajectories of speed, gap, and relative speed over 1000 seconds. Each curve corresponds to a vehicle, the color indicating its position in the platoon. The leader initiates a smooth trapezoidal acceleration-deceleration maneuver, and the system’s response is observed across the following vehicles. \Cref{time-space} presents the corresponding time–space diagram, where the vehicle position is plotted against time and colored by instantaneous speed.
Following the leader’s maneuver, a disturbance propagates through the follower vehicles. While oscillations emerge in the speed and gap trajectories, their amplitude does not exhibit significant growth along the platoon. In particular, relative speed fluctuations remain bounded across vehicles, and the time–space diagram shows that the speed wave does not intensify as it travels downstream. These observations suggest that the model exhibits marginal string stability under the given scenario. It is important to note that our primary objective is not to guarantee string stability through controller design. Rather, the focus of this work is to develop a data-driven framework that captures stochastic car-following behaviors with realistic temporal patterns. Nonetheless, the observed results indicate that the proposed model maintains acceptable stability properties in multi-vehicle simulations, further supporting its applicability to large-scale traffic modeling.

\begin{figure*}[!th]
    \centering
    \subfigure[Car-following Pair \#1.]{
        \centering
        \includegraphics[width=0.31\textwidth]{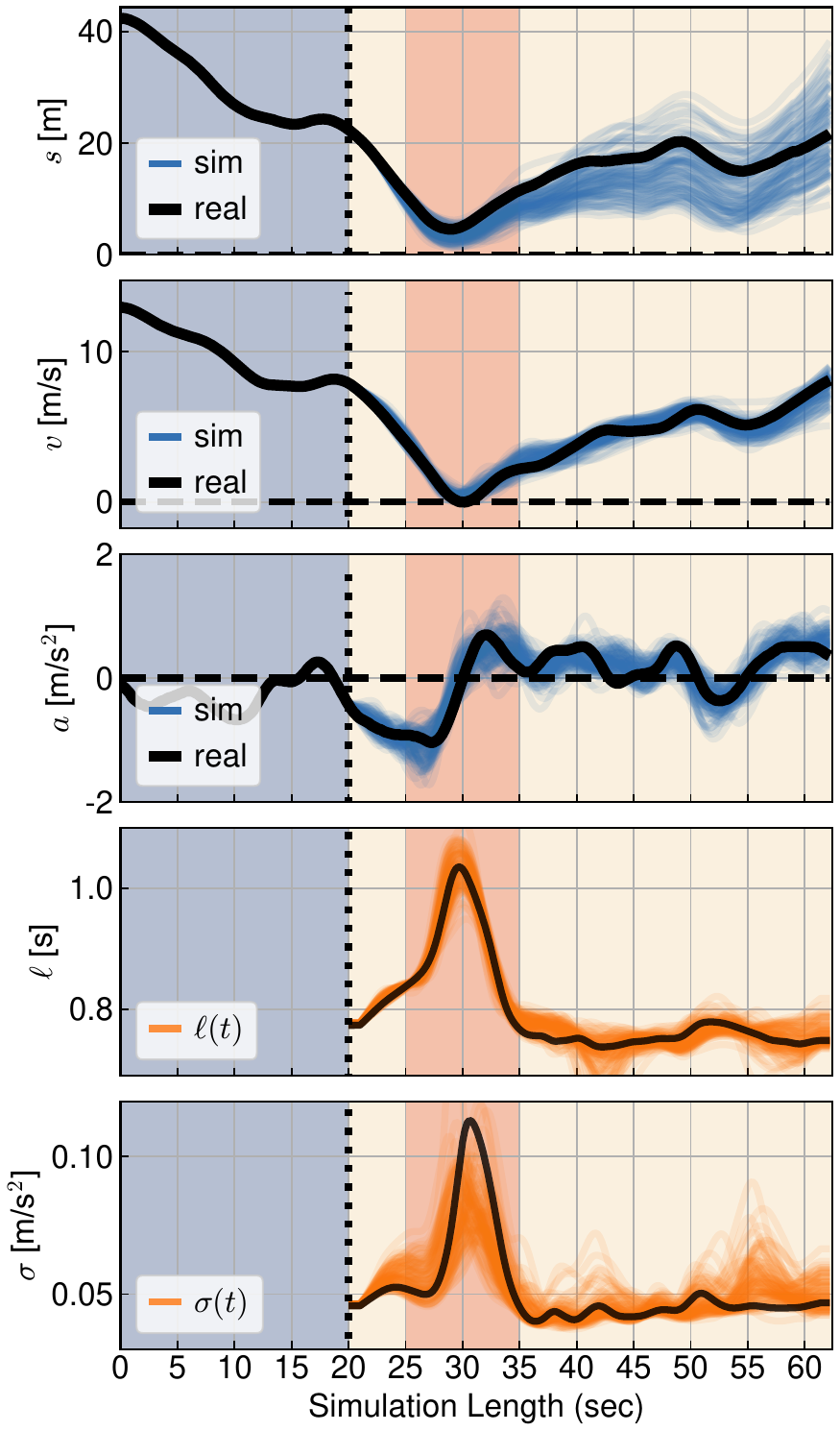}\label{S_1}
    }
    \subfigure[Car-following Pair \#2.]{
        \centering
        \includegraphics[width=0.305\textwidth]{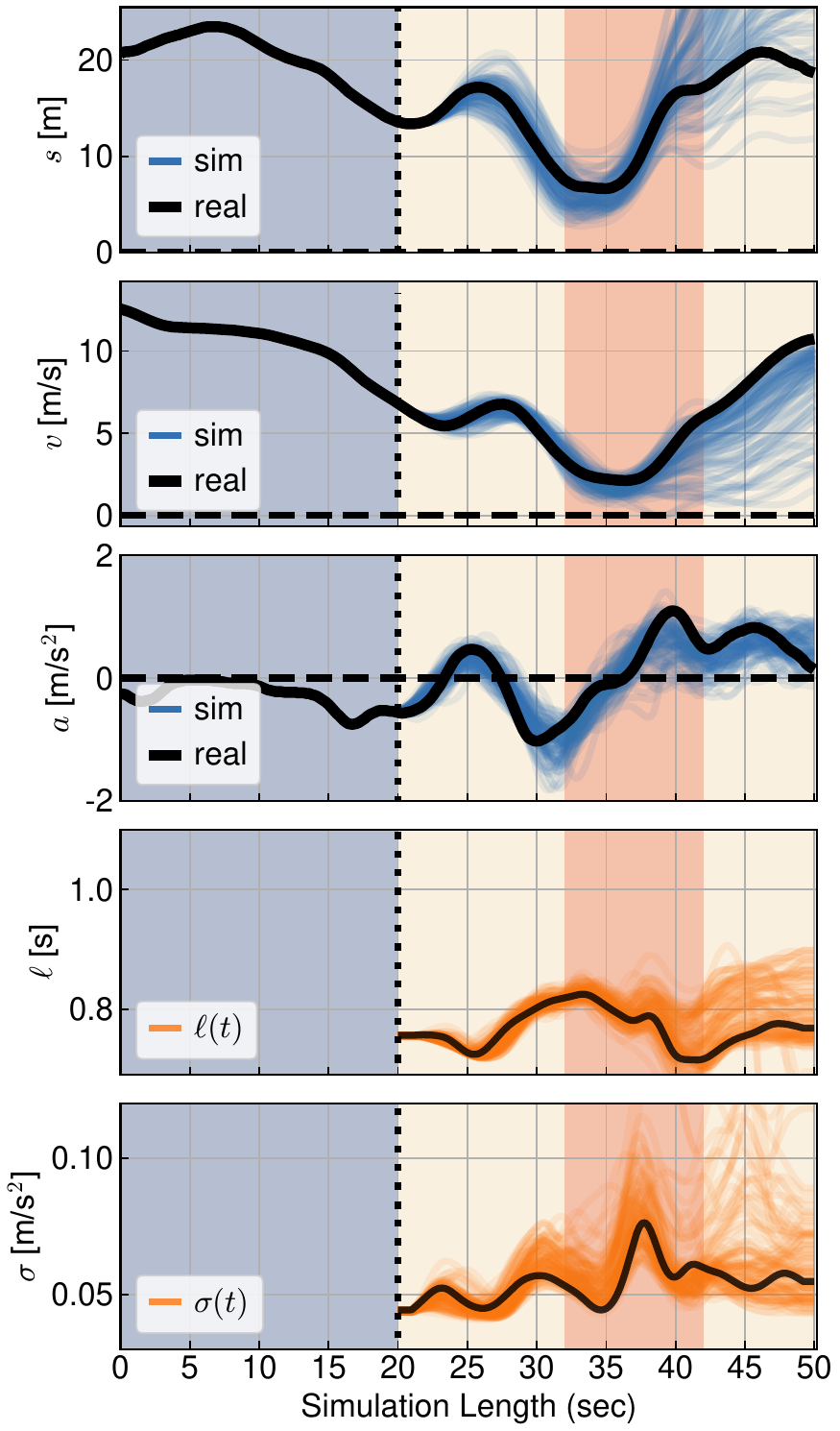}\label{S_2}
    }
    \subfigure[Car-following Pair \#3.]{
        \centering
        \includegraphics[width=0.305\textwidth]{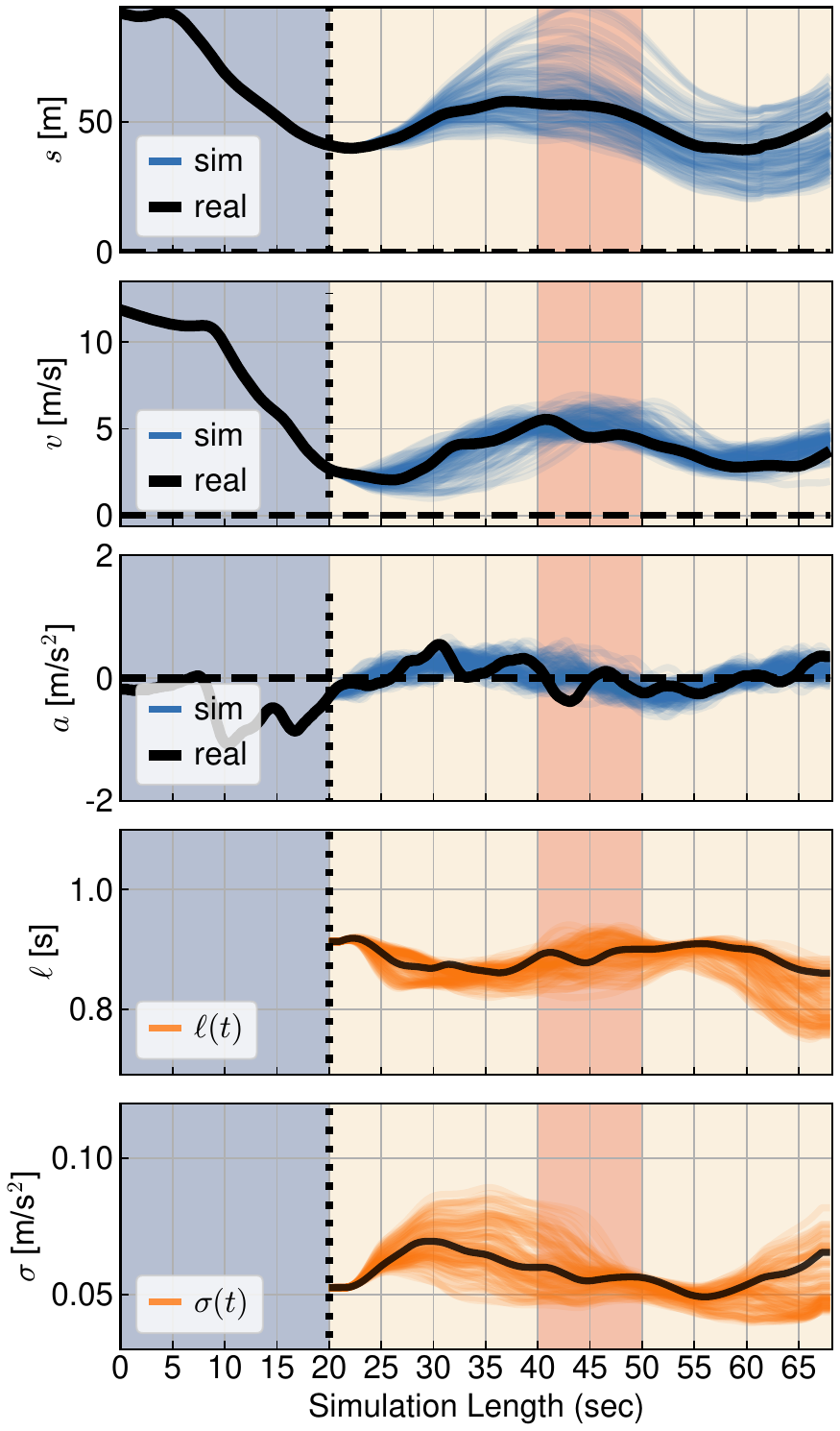}\label{S_3}
    }
    \subfigure[Example $\boldsymbol{K}$ in Pair \#1.]{
        \centering
        \includegraphics[width=0.31\textwidth]{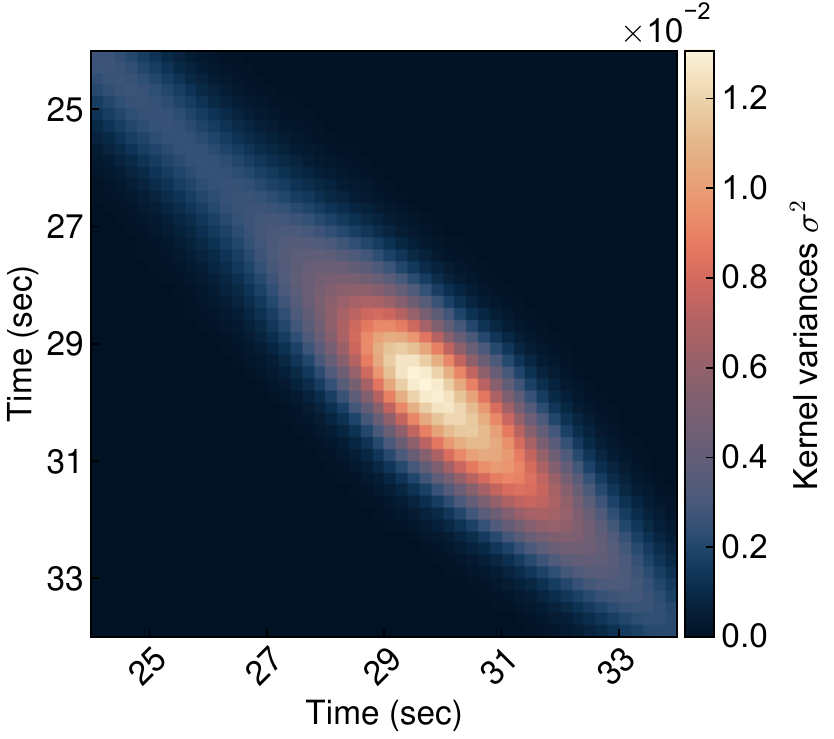}\label{K_1}
    }
    \subfigure[Example $\boldsymbol{K}$ in Pair \#2.]{
        \centering
        \includegraphics[width=0.30\textwidth]{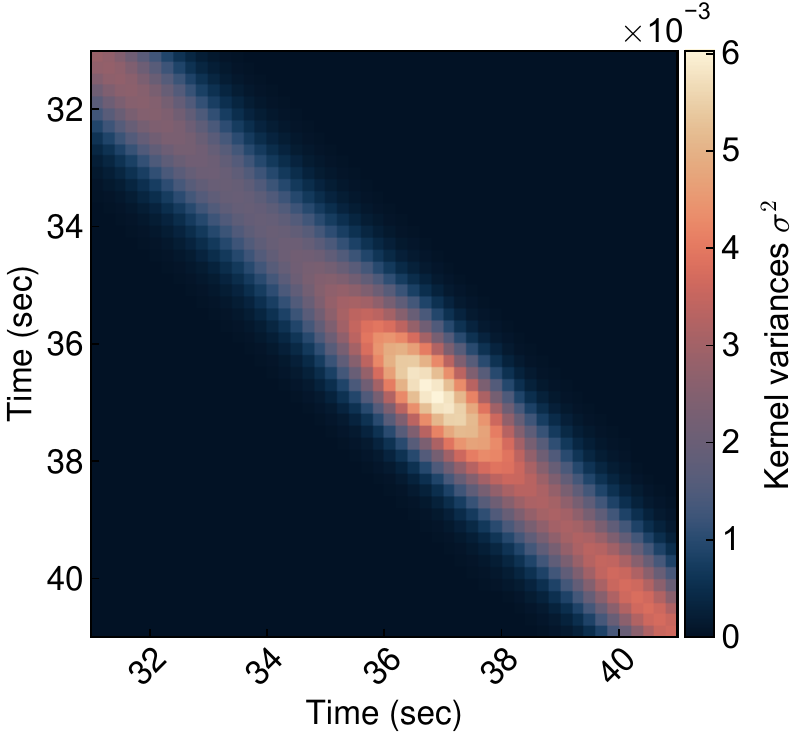}\label{K_2}
    }
    \subfigure[Example $\boldsymbol{K}$ in Pair \#3.]{
        \centering
        \includegraphics[width=0.31\textwidth]{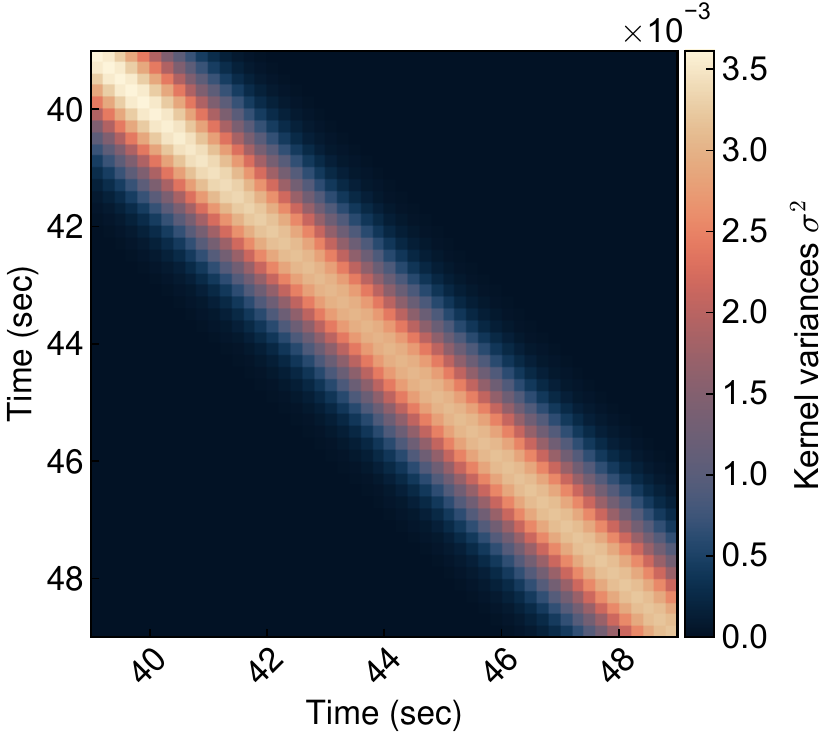}\label{K_3}
    }
    \caption{Stochastic simulation results for three representative car-following cases. (a)-(c): Each example shows 100 predicted trajectories of spacing $s$, speed $v$, and acceleration $a$ (blue), compared to the ground-truth (black). The dashed line marks the forecast start. Below each example, simulated context-dependent $\ell(\boldsymbol{x}_t)$ and $\sigma(\boldsymbol{x}_t)$ (orange) are compared to DeepAR conditioned on ground truth (black). (d)-(f): Bottom-row heatmaps visualize the GP kernel $\boldsymbol{K}$, revealing the evolving temporal correlation structure during the forecast horizon.}
    \label{fig:3_demos}
\end{figure*}

\subsubsection{Scenario-specific analysis and interpretability}
To demonstrate the effectiveness and interpretability of the proposed stochastic simulation framework, we present results generated by a trained DeepAR model equipped with a nonstationary Gibbs kernel. This model leverages context-dependent lengthscale $\ell(\boldsymbol{x}_t)$ and variance $\sigma^2(\boldsymbol{x}_t)$ to adapt dynamically to varying traffic scenarios, enabling the simulation of both steady and highly variable driving behaviors.

\Cref{S_1}-\Cref{S_3} illustrates 100 stochastic simulation rollouts for three representative car-following pairs. In each example, the top three rows display the simulated trajectories of gap $s$, speed $v$, and acceleration $a$. The black lines denote the ground-truth observations, while the blue lines represent sampled simulations. The vertical dashed line at $t=t_{\text{start}}$ separates the \textit{historical context window} (used for initialization) from the \textit{forecast region}, where the model generates future predictions. As shown, the simulated trajectories closely track the real motions, validating the model's ability to reproduce human-like car-following behaviors under uncertainty. The bottom two rows of each example plot show the time-varying $\ell(\boldsymbol{x}_t)$ and $\sigma(\boldsymbol{x}_t)$ values during the forecast horizon. The orange lines represent the model's predictions under simulation, while the black lines depict the outputs of DeepAR when conditioned on the true (non-simulated) trajectory. This dual presentation facilitates case-specific analysis and offers interpretability into the evolving behavioral regime.

To further illustrate the temporal structure learned by the model, we extract a 10-second simulation window from each forecast region and visualize the corresponding kernel matrix $\boldsymbol{K}$ as a heatmap in \Cref{K_1}-\Cref{K_3}. The diagonal entries of $\boldsymbol{K}$ reflect self-similarity and appear as the brightest band. The bandwidth around the diagonal conveys the extent of temporal correlation: Wider bands correspond to larger $\ell(\boldsymbol{x}_t)$, indicative of smoother and more persistent behavioral states; Narrower bands imply smaller $\ell(\boldsymbol{x}_t)$, capturing localized, rapidly evolving dynamics. The lighter regions (off-diagonal) signify stronger temporal correlations between points that are farther apart in time, often occurring in scenarios with a larger $\ell(\boldsymbol{x}_t)$ or smoother dynamics. In contrast, the darker regions represent weaker correlations, where the influence between time points diminishes due to either larger temporal distances or smaller $\ell(\boldsymbol{x}_t)$. We now discuss the behavioral interpretations of the three representative pairs:
\begin{itemize}
    \item Pair \#1, transitional scenario with braking and free acceleration: A sudden deceleration occurs just before 27 seconds, followed by a smooth acceleration phase. The inferred lengthscale $\ell(\boldsymbol{x}_t)$ increases steadily, peaking at around 30 seconds, which corresponds to the onset of free-flowing, anticipatory driving. This reflects the driver's transition from reactive braking to stable cruising. Concurrently, the variance $\sigma^2(\boldsymbol{x}_t)$ increases, capturing the rising behavioral heterogeneity\footnote{Driving heterogeneity refers to the diversity in driving styles and decision-making processes among different drivers. For instance, some drivers may brake or accelerate more aggressively, while others may respond more conservatively or gradually.}: though the situation is stable, drivers have the freedom to vary their responses (e.g., whether to follow closely or accelerate assertively). This interpretable separation of temporal correlation and uncertainty tolerance highlights the model’s ability to disentangle different driving regimes.
    \item Pair \#2, highly dynamic scenario with frequent mode switches: In this case, the driver alternates between acceleration and deceleration, requiring constant monitoring of the leading vehicle, road conditions, and other contextual factors (e.g., traffic flow). The learned $\ell(\boldsymbol{x}_t)$ remains low throughout, indicating that the behavior is rapidly changing and thus memory should be short. The variance $\sigma^2(\boldsymbol{x}_t)$ also remains small, reflecting a constrained decision space with limited tolerance for uncertainty, which is appropriate for safety-critical, high-attention driving episodes.
    \item Pair \#3, steady-state following behavior: The simulation reflects a consistently smooth driving pattern. Both $\ell(\boldsymbol{x}_t)$ and $\sigma^2(\boldsymbol{x}_t)$ exhibit moderate and stable values, indicating a scenario with infrequent decisions and minimal need for rapid response. This case resembles relaxed car-following where both the driver and model rely on slowly evolving, stable behavioral rules.
\end{itemize}

The qualitative differences across the three examples further validate the interpretability of the model’s internal representations. In \Cref{K_3},  the smooth and stationary structure of $\boldsymbol{K}$ aligns with the behavior expected under a stationary kernel such as the squared exponential (SE). However, in Pairs \#1 and \#2 (\Cref{K_1}-\Cref{K_2}), the nonstationary Gibbs kernel offers clear advantages. It captures dynamic adjustments in both $\ell(\boldsymbol{x}_t)$ and $\sigma^2(\boldsymbol{x}_t)$, providing:
\begin{enumerate}
    \item \textbf{A reactive lengthscale that contracts during rapid regime transitions and expands during smooth driving phases;}
    \item \textbf{A variance term that reflects behavioral heterogeneity---rising in permissive conditions and falling in tightly constrained, safety-critical contexts.}
\end{enumerate}

This flexibility is essential for modeling the stochastic, nonstationary nature of human driving, which deterministic frameworks cannot fully represent.


\subsection{Quantitative Evaluations of Simulation Performance}
To thoroughly evaluate both the accuracy and the uncertainty quantification performance of the proposed stochastic simulation framework, we adopt three complementary metrics: Root Mean Square Error (RMSE), Continuously Ranked Probability Score (CRPS), and Energy Score (ES). Each metric offers a distinct perspective on simulation fidelity, from point prediction accuracy to probabilistic calibration and multivariate distributional consistency.

RMSE quantifies the average deviation between the simulated and ground-truth trajectories for acceleration, speed, and gap distance, thus serving as a standard measure of prediction accuracy \citep{zhou2025calibration}. Formally, it is computed as:
\begin{equation}
    \text{RMSE}(\boldsymbol{x}) := \sqrt{\frac{1}{n} \sum_{t=1}^{n} \left( x_t - \hat{x}_t \right)^2},
\end{equation}
where $x_t$ and $\hat{x}_t$ denote the ground truth and simulated values, respectively, at time $t$, and $n$ the total number of time steps. Lower RMSE values indicate better alignment with empirical driving behaviors.

CRPS is employed to evaluate the quality of uncertainty quantification in one-dimensional stochastic forecasts. Unlike RMSE, which considers only the predicted mean, CRPS assesses the entire forecast distribution $F(x)$ against the observed outcome $x_t$, measuring both calibration and sharpness \citep{matheson1976scoring}:
\begin{equation}
    \mathrm{CRPS}(x_t) := \int_{-\infty}^{+\infty}\left(F(x)-\mathbbm{1}\{x>x_t\}\right)^2 dx,
\end{equation}
where $\mathbbm{1}\{\cdot\}$ is the indicator function. Lower CRPS values indicate better alignment of the predicted distributions with the observed data, reflecting the model's ability to provide reliable probabilistic forecasts.

ES extends the assessment to a multivariate setting, capturing how well the model reproduces the joint distribution of acceleration, speed, and spacing across multiple time steps. This is crucial for evaluating the coherence of simulated trajectories in multi-step forecasting. Formally, ES is defined as:
\begin{equation}
\text{ES}(\mathcal{P}, \boldsymbol{x}) := \frac{1}{M} \sum_{m=1}^M \lVert \boldsymbol{X}^{(m)} - \boldsymbol{x} \rVert - \frac{1}{2M^2} \sum_{m=1}^M \sum_{k=1}^M \lVert \boldsymbol{X}^{(m)} - \boldsymbol{X}^{(k)} \rVert, 
\end{equation}
where $\mathcal{P}$ is the predictive distribution, $\boldsymbol{x}$ is the observed vector of target variables, $\boldsymbol{X}^{(m)}$ are samples drawn from $\mathcal{P}$, $M$ is the number of samples, and $\lVert \cdot \rVert$ denotes the Euclidean norm. The ES penalizes discrepancies between the simulated samples and the ground truth while also penalizing inconsistencies within the simulated samples. This dual focus makes it an effective metric for evaluating joint predictive performance in multi-step stochastic simulations. Lower ES indicates better agreement between the predicted and observed multivariate distributions, reflecting the model’s ability to generate realistic and reliable forecasts.

To benchmark the simulation quality across different kernel designs, we conducted experiments on 30 car-following pairs selected from the held-out test set $\mathcal{D}_{\text{test}}$. For each test pair, we conducted 200 stochastic simulations following the procedure outlined in \Cref{sim_1_round}, with forecasts initiated at various start times $t_{\text{start}}$, uniformly sampled every 5 seconds along each trajectory. Simulations were run for a fixed horizon of 10 seconds (i.e., $t_{\text{end}}=t_{\text{start}}+10\,\text{s}$).

Each simulation was executed at a temporal resolution of 5 Hz, consistent with $\mathcal{D}_{\text{HighD}}$ dataset’s downsampled frame rate ($\Delta t=0.2$ s).  This resolution strikes a balance between capturing meaningful behavioral variation and maintaining computational efficiency. RMSE, CRPS, and ES were computed for each simulation round, aggregated across acceleration, speed, and spacing dimensions. This setup enables a rigorous evaluation of both pointwise trajectory accuracy and probabilistic consistency under various kernel configurations, including the Gibbs kernel (nonstationary), SE and Mat\'ern kernels (stationary), and a white noise baseline that assumes \textit{i.i.d.} residuals.

In the following section, we analyze the results of this comparative evaluation, highlighting how different kernel assumptions impact simulation fidelity and uncertainty quantification.

\subsubsection{Impact of kernel design on simulation performance}

\begin{table}[t]
    \footnotesize
    \centering
    \caption{Evaluations of 10-second stochastic simulations.}
    \resizebox{\textwidth}{!}{
    \begin{tabular}{c|c|c|c}
    \toprule
    Metrics $\backslash$ Model & DeepAR & MLP (with delay embedding) & Bayesian IDM (pooled)\\
    \hline
    RMSE($a$) & \textbf{0.215} / 0.216 / 0.218 / \textcolor{gray}{0.709} & \textbf{0.214} / 0.238 / 0.261 / \textcolor{gray}{0.529} & \textbf{0.273} / 0.275 / 0.277 / \textcolor{gray}{0.565}  \\
    RMSE($v$) & \textbf{0.429} / 0.430 / 0.434 / \textcolor{gray}{1.416} & \textbf{0.461} / 0.492 / 0.539 / \textcolor{gray}{1.088} & \textbf{0.711} / 0.718 / 0.722 / \textcolor{gray}{0.832}  \\
    RMSE($s$) & \textbf{1.380} / 1.384 / 1.414 / \textcolor{gray}{4.498} & \textbf{1.577} / 1.579 / 1.826 / \textcolor{gray}{3.625} & \textbf{2.420} / 2.423 / 2.455 / \textcolor{gray}{3.087}  \\
    CRPS($a$) & \textbf{0.146} / 0.146 / 0.149 / \textcolor{gray}{0.258} & \textbf{0.129} / 0.134 / 0.160 / \textcolor{gray}{0.195} & \textbf{0.216} / 0.220 / 0.221 / \textcolor{gray}{0.231}  \\
    CRPS($v$) & \textbf{0.259} / 0.264 / 0.266 / \textcolor{gray}{0.731} & \textbf{0.324} / 0.352 / 0.372 / \textcolor{gray}{0.448} & \textbf{0.772} / 0.804 / 0.809 / \textcolor{gray}{0.835}  \\
    CRPS($s$) & \textbf{1.305} / 1.354 / 1.362 / \textcolor{gray}{2.828} & \textbf{1.715} / 1.869 / 1.870 / \textcolor{gray}{2.472} & \textbf{3.640} / 3.864 / 3.871 / \textcolor{gray}{4.455}  \\
    ES($a$) & \textbf{0.911} / 0.933 / 0.939 / \textcolor{gray}{1.676} & \textbf{0.953} / 0.987 / 1.083 / \textcolor{gray}{1.385} & \textbf{1.463} / 1.465 / 1.485 / \textcolor{gray}{1.631}  \\
    ES($v$) & \textbf{1.748} / 1.764 / 1.781 / \textcolor{gray}{3.647} & \textbf{2.077} / 2.144 / 2.253 / \textcolor{gray}{3.011} & \textbf{3.854} / 4.003 / 4.030 / \textcolor{gray}{4.386}  \\
    ES($s$) & \textbf{5.572} / 5.638 / 5.671 / \textcolor{gray}{11.665} & \textbf{7.094} / 7.314 / 7.454 / \textcolor{gray}{10.248} & \textbf{13.078} / 13.756 / 13.799 / \textcolor{gray}{16.861}  \\
    \bottomrule
    \multicolumn{4}{p{.9\linewidth}}{\raggedright * The metrics are with the order of `\textbf{Gibbs}/Mat\'ern(5/2)/SE/\textcolor{gray}{White Noise}' kernel, corresponding to the models with `\textbf{nonstationary GP}/stationary GP/stationary GP/\textcolor{gray}{i.i.d.}' assumptions, respectively.}
    \end{tabular}}
    \label{tab:models_comparison}
\end{table}

\Cref{tab:models_comparison} summarizes the RMSE, CRPS, and Energy Score (ES) across acceleration, speed, and spacing for 10‐second stochastic simulations under four covariance structures: Gibbs kernel (nonstationary GP), Mat\'ern(5/2) kernel (stationary GP), SE kernel (stationary GP), and White Noise kernel (i.i.d.). { Compared to the simple white-noise baseline, the nonstationary Gibbs residual yields large absolute gains, while adding only a short-horizon covariance structure in simulation.} These results offer a comprehensive comparison of how kernel design influences both predictive accuracy and uncertainty quantification in car-following simulations.

Among all alternatives, the Gibbs kernel consistently achieves the lowest values across all three motion states and metrics. This strong performance underscores the advantages of a nonstationary formulation. Specifically, the Gibbs kernel’s scenario-varying lengthscale allows the model to adapt its temporal memory to local traffic conditions. This dynamic adjustment is particularly critical in capturing the nuanced behavior of drivers during rapid accelerations, abrupt decelerations, and transitions between free-flow and congested regimes. By tailoring the kernel width to the context, the Gibbs kernel not only improves predictive accuracy but also yields sharper and more reliable uncertainty estimates, especially in behaviorally complex or safety-critical episodes.

In contrast, the stationary kernels, while still capturing general temporal dependencies, exhibit reduced flexibility due to their fixed lengthscales. The Mat\'{e}rn kernel outperforms the SE kernel across most metrics, attributable to its additional smoothness parameter $\nu$, which moderates the rate of decay in correlations and permits more adaptable representations of moderate temporal dynamics. However, both stationary models struggle with the rapid variability characteristic of human driving behaviors, leading to higher residual errors and miscalibrated uncertainty bounds.

{ The White Noise model serves as a baseline representing the standard independent error assumption commonly found in traditional calibration studies \citep{hoogendoorn2010calibration, punzo2021calibration}.
By modeling each time step as conditionally independent (i.e., assuming \textit{i.i.d.} residuals), this configuration reflects the performance of models that neglect structured temporal dependencies.
As expected, it yields the highest RMSE, CRPS, and ES values across all states, demonstrating that standard calibration practices may lead to degraded performance in sequential tasks such as trajectory forecasting due to the failure to model temporal correlations.}

\begin{figure}[!t]
    \centering
    \includegraphics[width=.97\linewidth]{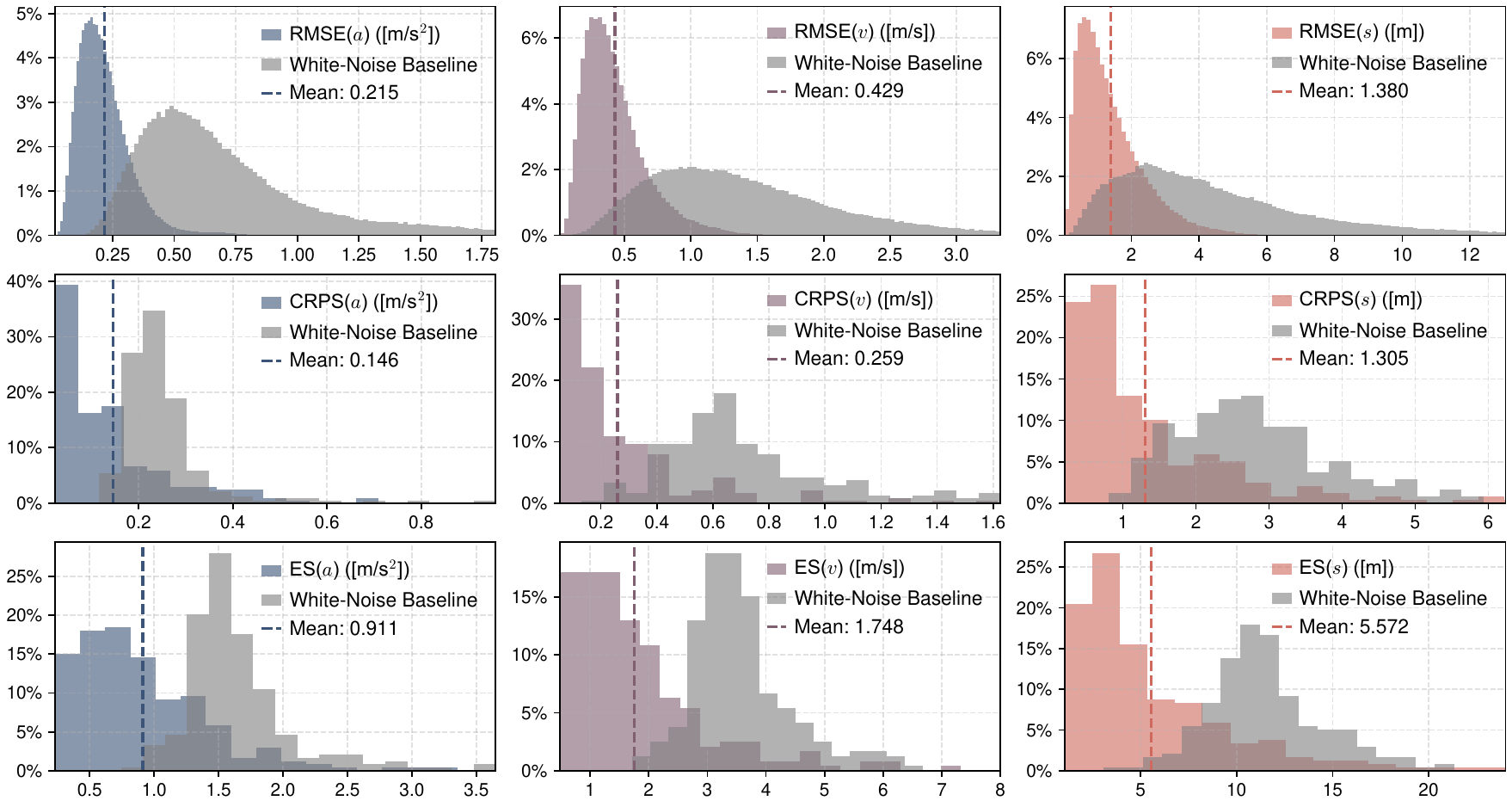}
    \caption{Error distributions under the DeepAR + Gibbs kernel (nonstationary GP) configuration over all test simulations. Each row corresponds to a different metric. Top: RMSE, middle: CRPS, bottom: ES. Each column corresponds to a different state (acceleration, speed, spacing). Histograms are normalized to percentage frequency; dashed vertical lines denote the mean values reported in \Cref{tab:models_comparison}. Note the long tails in the spacing histograms, indicating that although the average gap error is low, extreme underestimations or overestimations can still occur.}
    \label{fig:error_dist}
\end{figure}

To complement these aggregate statistics, \Cref{fig:error_dist} presents the full distribution of RMSE, CRPS, and ES across all test trajectories for each motion state. These histograms reveal the dispersion and tail behavior of errors that are not captured by mean scores alone. Notably, for the spacing variable, even the best-performing models occasionally produce large deviations, highlighting edge cases where the simulation diverges from ground truth. Such worst-case discrepancies are crucial for understanding the practical limits of each model’s reliability, especially in real-world applications where safety is a concern. Together, \Cref{tab:models_comparison} and \Cref{fig:error_dist} emphasize the importance of incorporating nonstationary temporal correlations for accurately modeling human-like driving behavior under uncertainty.

\subsubsection{Comparative model analysis and temporal correlation effects}
To assess the generalizability and robustness of our proposed DeepAR-Gibbs framework, we benchmark its performance against alternative neural architectures and a classical model. Specifically, we evaluate: (i) a multilayer perceptron (MLP) with delay embedding, (ii) the Bayesian IDM, and (iii) ablated versions of each model that assume temporally uncorrelated (white noise) residuals.

Across all evaluation metrics, DeepAR combined with the nonstationary Gibbs kernel exhibits superior performance. Notably, it achieves the lowest RMSE across the three covariates, indicating its effectiveness in capturing abrupt and context-dependent changes in driver behavior. In terms of uncertainty quantification, the CRPS values are also consistently lower than those produced by models using stationary kernels, validating the Gibbs kernel’s capacity to adaptively characterize residual variability. The ES further confirms this advantage, all values substantially outperform the corresponding metrics from models employing Mat\'ern or SE kernels.

When comparing DeepAR to the MLP baseline (also using a Gibbs kernel), the recurrent structure of DeepAR provides measurable improvements. Although MLP achieves a marginally better CRPS on acceleration, its RMSE and ES on spacing are notably higher, suggesting a diminished ability to model long-range temporal dependencies. The degradation in performance becomes more pronounced when stationary kernels are used for either architecture, and models assuming white noise residuals perform the worst across all metrics. These results underscore the importance of integrating nonstationary temporal structures into residual modeling to improve both fidelity and uncertainty calibration.

The classical Bayesian IDM (pooled) performs the worst among all models, with substantially higher RMSE, CRPS, and ES values. This highlights the limitations of rule-based formulations in reproducing the stochastic variability observed in human driving. However, when the IDM is paired with a nonstationary Gibbs kernel, its performance improves markedly. Surprisingly, it even outperforms DeepAR with white noise residuals in several metrics. This result emphasizes a key insight: modeling structured residuals via nonstationary kernels can outweigh the benefits of using more complex architectures without such structure. Therefore, \textbf{even simple models can outperform sophisticated neural networks when temporal correlations are properly captured}.

This insight is further reinforced by an ablation study, in which the kernel matrix $\boldsymbol{K}$ is replaced by a diagonal matrix $\sigma^2\boldsymbol{I}_{\Delta T}$, effectively assuming \textit{i.i.d.} noise. The performance drop is significant across all models and metrics. The absence of temporal correlations particularly affects dynamic scenarios like sudden braking or acceleration, where time-dependent behavior is crucial. Additionally, uncertainty estimates become miscalibrated, which is overconfident during transitions and overly conservative in stable regimes.

In summary, these results reinforce the following conclusions: 
\begin{enumerate}
    \item Nonstationary kernels (Gibbs) yield the most accurate and reliable results across all model types, demonstrating that they are essential for capturing the evolving temporal structure in car-following behavior;
    \item DeepAR outperforms MLP due to its recurrent design, especially when modeling long-term dependencies;
    \item Bayesian IDM is outperformed by the proposed deep-learning-based method, though its performance improves significantly with nonstationary residual modeling;
    \item \textbf{Temporal correlation modeling is not just a refinement but a necessity} for generating realistic and interpretable stochastic driving simulations, regardless of model complexity.
\end{enumerate}

{
\subsection{Discussion: Framework Synthesis, Practical Implications, and Transferability}
\label{sec:discussion}

\paragraph{Framework synthesis.}
This paper argues that a substantial portion of car-following variability remains unexplained even after conditioning on commonly used contextual variables, and that this residual exhibits temporally structured dependence that can change across driving regimes. Consistent with this view, replacing an \textit{i.i.d.}\ disturbance with a temporally correlated residual improves both predictive accuracy and probabilistic consistency, with the largest gains when correlation structure is allowed to vary over time. These findings support a coupled mean--residual decomposition: a mean dynamics captures predictable interaction structure, while a stochastic residual process captures regime-dependent, temporally correlated deviations beyond the mean. The benefit arises from their synergy: improving the mean reduces systematic bias, whereas modeling residual dependence improves uncertainty representation and temporal realism.

\paragraph{Model-agnostic residual module and kernel choice.}
Although we instantiate the mean dynamics using DeepAR, the nonstationary GP is defined on residuals and can be combined with a broad family of car-following models, including more behaviorally structured formulations (e.g., safety-constrained, kinematic, traffic-flow-inspired, and psycho-physical models). The kernel choice should reflect residual characteristics: a stationary SE kernel can suffice when residuals are smooth and stable within an episode; Mat\'ern kernels are preferable for rougher or faster-varying residuals; and the nonstationary Gibbs kernel is most useful when correlation lengthscales and variance shift across regimes (e.g., transitions from free-flow to close following), where a single global lengthscale is inadequate.

\paragraph{Transferability under domain shift.}
Because the framework learns both mean dynamics and context-dependent residual structure from data, performance depends on how well deployment conditions are represented by the training distribution. Under substantial domain shift (e.g., different road geometry, traffic rules, or driver populations), generalization is not guaranteed. A useful separation is what may transfer (shared regimes and stable physical meaning of context variables such as headway and relative speed) versus what may require adaptation (qualitatively different interaction patterns). In practice, lightweight adaptation can be implemented by fine-tuning a subset of parameters using a small target-domain calibration set, for example updating kernel modulation and output scales, or adapting the mean while retaining kernel structure when temporal dependence appears stable.

\paragraph{Measurement noise and evaluation.}
Acceleration is often obtained by differentiating discrete trajectories, which can amplify measurement noise and distort residual autocorrelation if not accounted for. Our formulation partially addresses this by separating observation noise from the temporally correlated residual: the additive noise term absorbs unstructured measurement error, while the GP residual represents structured, lower-frequency deviations beyond the mean. In practice, preprocessing (e.g., smoothing or downsampling prior to differentiation) and plausibility checks (e.g., jerk statistics) are important; a principled extension is to treat acceleration as latent and introduce an explicit measurement model. Finally, evaluation should match downstream use: probabilistic scores such as CRPS complement point metrics for forecasting, while simulation requires additional checks of physical plausibility, stability, and reproduction of emergent stop-and-go phenomena.

\paragraph{Identifiability and noise decomposition.}
Following the above discussion, a critical aspect of the proposed framework is the decomposition of the residual into a temporally correlated behavioral process $\delta(t)$ and an i.i.d.\ observation noise $\epsilon$. This separation is essential for model identifiability; mathematically, if both components were permitted to exhibit temporal correlation, the likelihood optimization would face a significant signal-separation challenge, struggling to distinguish between structured driver behavior and high-frequency measurement artifacts. By treating $\epsilon$ as a ``nugget effect,'' we absorb irreducible sensor noise and differentiation errors into the white-noise term. This formulation forces the GP kernel to prioritize the capture of lower-frequency, persistent behavioral dependencies rather than measurement artifacts. The same identifiability issue of the Bayesian method within this framework was previously discussed in~\citet{zhang2024bayesian} and~\citet{zhang2024calibrating}.

\paragraph{Practical deployment and evaluation in microsimulation.}
While our experiments validate short-horizon stochastic rollouts on naturalistic trajectories, an important next step is closed-loop microsimulation evaluation, where the model is used as the driving policy for multiple interacting vehicles and assessed via macroscopic outcomes (e.g., wave propagation, stop-and-go formation, and safety-related events). Moreover, broader cross-dataset validation is needed to quantify robustness under heterogeneous road types, regions, and driver populations. For simulator integration, a practical route is to use the selected mean model as the nominal car-following rule and inject a temporally correlated residual in a receding-horizon manner (apply the first-step acceleration each tick), enabling plug-in deployment without changing the simulator’s underlying vehicle dynamics.

\paragraph{Extensions to delayed, non-smooth, and multi-lane settings.}
The residual-based construction provides a direct path to richer settings. For delayed-response models (e.g., GM-type), the mean can incorporate delayed covariates or an explicit reaction-time parameter while retaining the GP to capture remaining temporally correlated variability. For action-point or piecewise models with non-smooth accelerations, e.g., Wiedemann-type \citep{wiedemann1974simulation}, the mean can encode discontinuous updates, and the residual module can accommodate rougher deviations via Mat\'ern or regime-dependent nonstationarity. Multi-lane extensions can be achieved by augmenting context with adjacent-lane states and maneuver cues (e.g., neighboring gaps, closing rates, lane-change indicators), allowing uncertainty to adapt during lane-changing and merging interactions.

}

\section{Conclusion}\label{sec:conclusion}

This work presents a stochastic simulation framework for interpretable modeling of car-following behaviors, centered on the integration of a nonstationary GP kernel with a neural predictive architecture. Leveraging DeepAR as the backbone, the proposed model learns complex, nonlinear driving dynamics from trajectory data while explicitly modeling structured residual uncertainty that persists beyond what observable context can explain. This distinction between context-driven predictions and unexplained variability enables simulations that are both behaviorally realistic and uncertainty-aware.

A key contribution of this framework is the use of a scenario-adaptive Gibbs kernel, which introduces context-dependent lengthscale and variance to model nonstationary temporal correlations in driver behavior. The learned lengthscale $\ell$ serves as \textit{a proxy for driver reaction frequency}, while the kernel variance $\sigma_k^2$ reflects \textit{the driver's tolerance for behavioral variability}. These parameters not only enhance the realism of stochastic simulations but also offer interpretable signals for identifying risky or adaptive driving patterns, of particular importance for safety-critical applications.

Empirical evaluations demonstrate the superiority of the proposed framework over conventional and baseline approaches. The Gibbs kernel consistently achieves lower RMSE, CRPS, and ES across all motion states, outperforming stationary alternatives such as the SE and Mat\'{e}rn(5/2) kernels. Furthermore, ablation studies confirm the essential role of temporal correlation modeling: removing the kernel structure in favor of \textit{i.i.d.} noise assumptions leads to substantial performance degradation in both accuracy and uncertainty quantification.

{
Overall, this study provides strong evidence that integrating structured, nonstationary residual modeling improves both simulation fidelity and uncertainty calibration for car-following. These capabilities support practical applications such as uncertainty-aware integration into microscopic traffic simulators and safety-critical connected and automated vehicles (CAV) analysis through risk-aware scenario evaluation and robustness testing under human-driving uncertainty.

}

\section*{Acknowledgement}
This research is supported by the Natural Sciences and Engineering Research Council of Canada (NSERC) of Canada. C. Zhang would also like to thank the McGill Engineering Doctoral Awards (MEDA), the Interuniversity Research Centre on Enterprise Networks, Logistics and Transportation (CIRRELT), and Fonds de recherche du Québec -- Nature et technologies (FRQNT) for providing scholarships and funding to support his Ph.D. study.

\bibliographystyle{apalike} 
\bibliography{reference}

\newpage
\section*{Appendix}
\subsection*{Comprehensive likelihood of multiple motion states}
It is more flexible to learn the model on multiple motion states, instead of only learning on acceleration data. According to the dynamic updating mechanism in \Cref{sim_1_round}, we could obtain the predicted motion states $\boldsymbol{v}^{\text{batch}}_{\text{NN}}$ and $\boldsymbol{p}^{\text{batch}}_{\text{NN}}$ based on $\boldsymbol{a}^{\text{batch}}_{\text{NN}}$. Therefore, we have their relationships as
\begin{align}
    \left[\begin{array}{c}
          \boldsymbol{a}^{\text{batch}} \\ \boldsymbol{v}^{\text{batch}} \\
          \boldsymbol{p}^{\text{batch}}
    \end{array}\right] =&  \left[\begin{array}{c}
          \boldsymbol{a}^{\text{batch}}_{\text{NN}} \\ \boldsymbol{v}^{\text{batch}}_{\text{NN}} \\
          \boldsymbol{p}^{\text{batch}}_{\text{NN}}
    \end{array}\right]  + \underbrace{\left[\begin{array}{c}
         1 \\
         \Delta t  \\
         \frac{1}{2}\Delta t^2
    \end{array}\right]}_{\coloneqq\boldsymbol{C}} \boldsymbol{\delta}_{\mathrm{GP}}+\left[\begin{array}{ccc}
          \sigma_{a}^2\boldsymbol{I}_{\Delta T} & \boldsymbol{0}_{T} & \boldsymbol{0}_{T} \\   \boldsymbol{0}_{T} &\sigma_{v}^2\boldsymbol{I}_{\Delta T}  & \boldsymbol{0}_{T}\\
         \boldsymbol{0}_{T} & \boldsymbol{0}_{T} & \sigma_{p}^2\boldsymbol{I}_{\Delta T}
    \end{array}\right],
\end{align}
where $\boldsymbol{0}_{T}\in\setR^{T\times T}$ denotes a matrix containing zeros and $\boldsymbol{I}_{T}\in\setR^{T\times T}$ denotes an identity matrix.
which is equivalent to
\begin{align}
    \left[\begin{array}{c}
          \boldsymbol{a}^{\text{batch}} \\ \boldsymbol{v}^{\text{batch}} \\
          \boldsymbol{p}^{\text{batch}}
    \end{array}\right]\,\sim\, \mathcal{N}\left(\left[\begin{array}{c}
          \boldsymbol{a}^{\text{batch}}_{\text{NN}} \\ \boldsymbol{v}^{\text{batch}}_{\text{NN}} \\
          \boldsymbol{p}^{\text{batch}}_{\text{NN}}
    \end{array}\right], \underbrace{\left[\begin{array}{ccc}
        1 & \Delta t & \frac{1}{2}\Delta t^2\\
        \Delta t & \Delta t^2 & \frac{1}{2}\Delta t^3 \\
        \frac{1}{2}\Delta t^2 &\frac{1}{2}\Delta t^3 &  \frac{1}{4}\Delta t^4
    \end{array}\right]}_{=\boldsymbol{C}\boldsymbol{C}^\top} \otimes \boldsymbol{K} + \underbrace{\left[\begin{array}{ccc}
        \sigma^2_a & 0 & 0 \\
        0 & \sigma^2_v & 0 \\
        0 & 0 & \sigma^2_p
    \end{array}\right]}_{:=\boldsymbol{\Sigma_{\text{obs}}}}\otimes\boldsymbol{I}_{\Delta T}\right),
\end{align}
where the covariance matrix incorporates both motion dynamics (via $\boldsymbol{C}\boldsymbol{C}^\top\in\setR^{3\times3}$) and temporal correlations (via $\boldsymbol{K}\in\setR^{T\times T}$), and we use $\otimes$ to denote the Kronecker product, which gives us a $3T\times 3T$ covariance matrix. $\sigma_a^2$, $\sigma_v^2$, and $\sigma_p^2$ are the variances of the observation noise for acceleration, speed, and position, respectively. A higher noise variance suggests less reliable data, contributing less to the likelihood.

Therefore, the log-likelihood function in \Cref{opt_problem} could be modified to incorporate more information as
\begin{align}
    \log \mathcal{L}
    &= -\frac{1}{2} \left( 3T \log (2\pi) + \log \det(\boldsymbol{\Sigma}) + \boldsymbol{r}^\top \boldsymbol{\Sigma}^{-1} \boldsymbol{r} \right),
\end{align}
where $\boldsymbol{\Sigma} := \boldsymbol{C} \boldsymbol{C}^\top \otimes \boldsymbol{K} + \boldsymbol{\Sigma}_{\text{obs}} \otimes \boldsymbol{I}_{\Delta T}$, and $\boldsymbol{r} := \begin{bmatrix}
        \boldsymbol{a}^{\text{batch}} \\ 
        \boldsymbol{v}^{\text{batch}} \\ 
        \boldsymbol{p}^{\text{batch}}
\end{bmatrix} - \begin{bmatrix}
    \boldsymbol{a}^{\text{batch}}_{\text{NN}} \\ 
    \boldsymbol{v}^{\text{batch}}_{\text{NN}} \\ 
    \boldsymbol{p}^{\text{batch}}_{\text{NN}}
\end{bmatrix}$.

\end{document}